%% file: gfamm_ejs.tex
\documentclass[ejs, preprint]{imsart}\usepackage[]{graphicx}\usepackage[]{color}
\makeatletter
\def\maxwidth{ %
  \ifdim\Gin@nat@width>\linewidth
    \linewidth
  \else
    \Gin@nat@width
  \fi
}
\makeatother

\definecolor{fgcolor}{rgb}{0.345, 0.345, 0.345}

\usepackage{framed}
\makeatletter
 {\par\unskip\endMakeFramed%
 \at@end@of@kframe}
\makeatother

\definecolor{shadecolor}{rgb}{.97, .97, .97}
\definecolor{messagecolor}{rgb}{0, 0, 0}
\definecolor{warningcolor}{rgb}{1, 0, 1}
\definecolor{errorcolor}{rgb}{1, 0, 0}
\newenvironment{knitrout}{}{} 

\usepackage{alltt}

\RequirePackage[numbers]{natbib}
\doi{10.1214/XXXXXXXXXXXXX}
\pubyear{0000}
\volume{0}
\firstpage{0}
\lastpage{0}

\newcommand{\backmatter}

\startlocaldefs
\input{header.tex}
\usepackage[modulo, displaymath, mathlines]{lineno}

\endlocaldefs
\IfFileExists{upquote.sty}{\usepackage{upquote}}{}

\IfFileExists{upquote.sty}{\usepackage{upquote}}{}
\begin{document}

\begin{frontmatter}

\title{Generalized functional additive mixed models}
\runtitle{GFAMM}

\author{\fnms{Fabian} \snm{Scheipl$^\star$}\corref{}\ead[label=e1]{[fabian.scheipl, sonja.greven]@stat.uni-muenchen.de}}
\and
\author{\fnms{Jan} \snm{Gertheiss$^\bullet$}\ead[label=e2]{jan.gertheiss@tu-clausthal.de}}
\and
\author{\fnms{Sonja} \snm{Greven$^\star$}}
\address{$\star$: Institut f{\"u}r Statistik \\ Ludwig-Maximillians-Universit{\"a}t M{\"u}nchen \\ Ludwigstrasse 33 \\ Munich, Germany \\\printead{e1}}
\address{$\bullet$:
Institut f{\"u}r Angewandte Stochastik und Operations Research
 \\ TU Clausthal \\ Erzstrasse 1 \\ Clausthal-Zellerfeld, Germany \\\printead{e2}}

\runauthor{Scheipl, Gertheiss, Greven}

\begin{abstract}
\input{gfamm_abstract.tex}
\end{abstract}




\end{frontmatter}



\input{gfamm_main.tex}


\bibliography{gfamm}

\clearpage

\input{gfamm_app.tex}

\end{document}

%% file: gfamm_abstract.tex
We propose a comprehensive framework for additive regression models for non-Gaussian functional responses, allowing for multiple (partially) nested or crossed functional random effects with flexible correlation structures for, e.g., spatial, temporal, or longitudinal functional data as well as linear and nonlinear effects of functional and scalar covariates that may vary smoothly over the index of the functional response. 
Our implementation handles functional responses from any exponential family distribution as well as many others like Beta- or scaled and shifted $t$-distributions.
Development is motivated by and evaluated on an application to large-scale longitudinal feeding records of pigs. Results in extensive simulation studies as well as replications of two previously published simulation studies for generalized functional mixed models demonstrate the good performance of our proposal. The approach is implemented in well-documented open source software in the \code{pffr} function in R-package \pkg{refund}.

%% file: gfamm_main.tex
\input{gfamm_intro.tex}

\input{gfamm_model.tex}
\input{gfamm_data.tex}


\input{gfamm_sim.tex}


\input{gfamm_discuss.tex}
 
\ifdefined \backmatter 
   \backmatter
\fi
 
\section*{Acknowledgements}
 
Sonja Greven and Fabian Scheipl were funded by Emmy Noether grant GR 3793/1-1 from the German Research Foundation. We appreciate Engel Hessel's permission to use the PIGWISE data collected in the framework of the ICT-AGRI era-net project PIGWISE ``Optimizing performance and welfare of fattening pigs using High Frequent Radio Frequency Identification (HF RFID) and synergistic control on individual level'' (Call for transnational research projects 2010). We are grateful for Bo Wang's patient and generous support for our attempt to reproduce the \code{ggpfr} results. We are indebted to the associate editor and three anonymous reviewers whose constructive remarks helped to improve the manuscript.

%% file: gfamm_intro.tex
\section{Introduction}

Data sets in which measurements consist of curves or images instead of scalars -- i.e., functional data -- are becoming ever more common in many areas of application. This is due to the increasing affordability and deployment of sensors like accelerometers or spectroscopes, high-throughput imaging technologies and automated logging equipment that continuously records conditions over time. Recent methodological development in this area has been rapid and intense, see \citet{Morris2015} for a review of the state of the art for regression models for functional data.

In this work, we extend the general framework for functional additive mixed models for potentially correlated functional Gaussian responses described in \citet{famm2014} to non-Gaussian functional responses. The development is motivated by and evaluated on an animal husbandry dataset in which the feeding behavior of growing-finishing pigs was monitored continuously over 3 months \citep{Maselyne2014, Gertheiss2014}. 
These are non-Gaussian functional data in the sense that the underlying probability
of feeding is assumed to be a continuous function over time, while the available data are sequences of binary indicators ("feeding: yes/no") evaluated at the temporal resolution of the sensors. Aggregating these binary indicators over given time intervals, we get time series of counts or proportions for which (truncated) Poisson, Negative Binomial, Binomial or Beta distributional assumptions could be appropriate. Another example of non-Gaussian functional data with large practical relevance would be continuously valued functional responses with heavy-tailed measurement errors for which a scaled $t$-distribution could be appropriate.

We briefly summarize the most relevant prior work on regression models for non-Gaussian functional responses. The fundamental work of \citet{Hall2008} relates observed functional binary or count data to a latent Gaussian process (GP) through a link function. This is the underlying idea of almost all the works that follow. They differ primarily in 1) how the latent functional processes are represented (i.e., either spline, wavelet or functional principal component (FPC) representations or full GP models), 2) to what extent additional covariate information can be included, 3) whether they allow the modelling of dependencies between and along the functional responses, 4) which distributions are available for the responses, 5) whether the functional data has to be available on a joint regular grid, and 6) in the availability of documented and performant software implementations.
In \citet{Hall2008}, the latent GP is represented in terms of its FPCs and neither correlated responses nor covariates are accommodated.  No implementation is publicly available. A Bayesian variant of \citet{Hall2008} is provided by \citet{vanDerLinde2009}. \citet{ZhuEtal2011} describe a robustified wavelet-based functional mixed model with scalar covariates for error-contaminated continuous functional responses on regular grids. No implementation was publicly available at the time of writing. \citet{SerbanStaicuCarroll2013} extend the approach of \citet{Hall2008} to  multilevel binary data without covariates and provide a rudimentary problem-specific implementation. \citet{WangShi2014} describe an empirical Bayesian approach for latent Gaussian process regression models for data from exponential families with scalar and concurrent functional covariates. The available implementation does not accommodate covariate effects and is limited to binary data with curve-specific $\iid$ functional random effects, see Section \ref{app:wang} for a systematic comparison with our proposal based on a replication of their simulation study.
\citet{Lietal2014} present a model for concurrent binary and continuous functional observations. Association between the two is modeled via cross-correlated (latent) FPC scores and cannot take into account any other covariate effects or dependency structures.
A similar approach is described in \citet[][ch. 3]{Tidemann2014}. \citet{Goldsmith2015}
develop a fully Bayesian approach which uses latent FPCs in a spline basis representation to represent multilevel functional random effects and linear functional effects of scalar covariates. They provide an implementation for binary outcomes with a logit link function, see Section \ref{app:goldsmith} for a systematic comparison with our proposal based on a replication of their simulation study.
\citet{Gertheiss2014} describe a marginal GEE-type approach for correlated binary functional responses with concurrent functional covariates. \citet{Brockhaus2015} develop a framework for estimating flexible functional regression models via boosting with similar flexibility as ours, implemented in R \citep{R} package \pkg{FDboost} \citep{fdboost}. The package additionally implements functional quantile regression models, which are not included in the framework we present here. Since their implementation is based on a component-wise gradient boosting algorithm, they cannot provide hypothesis tests or resampling-free construction of confidence intervals, however, and also have to rely on computationally intensive resampling methods for hyperparameter tuning. 

Compared to previous work, the novel contribution of this work is the development, implementation and evaluation of a comprehensive maximum likelihood-based inferential framework for generalized functional additive mixed models (GFAMMs) for potentially correlated functional responses. Our proposal accommodates diverse latent-scale correlation structures as well as flexible modeling of the conditional mean structure with multiple linear and non-linear effects of both scalar and functional covariates and their interactions. Our proposal is implemented in the full generality described here for both regular grid data and sparse or irregularly observed functional responses in the \code{pffr} function in R package \pkg{refund} \citep{refund}. Available response distributions include all exponential family distributions as well as Beta, scaled and shifted t-, Negative Binomial, Tweedie and zero-inflated Poisson distributions, each with various link functions, as well as cumulative threshold models for ordered categorical responses. With the exception of \citet{Brockhaus2015}, none of the previous proposals in this area achieve anything close to this level of generality, not to mention offer publicly available, widely applicable open-source implementations.
Since our framework is a natural extension of previous work done on generalized additive mixed models for scalar data and builds on the high performing, flexible implementations available for them, we can directly make use of many results from this literature, such as improved confidence intervals and tests for smooth effects \citep{MarraWood2012, Wood2013}.

The remainder of this work is structured as follows: Section \ref{sec:model} introduces notation and the theoretical and inferential framework for our model class. Section \ref{sec:app} presents results for our application. Section \ref{sec:sim} summarizes results of our extensive validation on synthetic data and of our partial replication of the simulation studies of \citet{WangShi2014} and \citet{Goldsmith2015}. Section \ref{sec:discuss} concludes.



%% file: gfamm_model.tex
\section{Model} \label{sec:model}

\subsection{Model Structure}
In what follows, we discuss structured additive regression models of the general form 
\begin{align}
\begin{split}
y_i(t) &\sim \mathcal{F}(\mu_i(t), \mnu)\\
g(\mu_i(t)) &= \eta_{i}(t) = \sum^R_{r=1} f_r(\mathcal{X}_{ri}, t),
\end{split}
\label{eq:fdmodel}
\end{align}
where, for each $t$, \mbox{$y_i(t), i=1,\dots,n,$} is a random variable 
from some distribution $\mathcal{F}$ with conditional expectation 
\mbox{$E(y_i(t)|\mathcal{X}_{i}, t, \mnu)  = \mu_i(t)$} observed over a domain 
$\mathcal{T}$ and an optional vector of nuisance parameters $\mnu$. We use 
$g(\cdot)$ to denote the known link function. Our implementation allows analysts
to choose $\mathcal{F}$ from the exponential family distributions as well as 
Tweedie, Negative Binomial, Beta, ordered categorical, zero-inflated Poisson and
scaled and shifted $t$-distributions. 
Note that, in the special case of ordered categorical responses, $\mu_i(t)$ is not the conditional mean of the response itself, but that of a latent variable whose value determines the response category.

Each term in the additive predictor is a function of a) the index $t$ of the 
response and b) a subset $\mathcal{X}_r$ of the complete covariate set 
$\mathcal{X}$ potentially including scalar and functional covariates and 
(partially) nested or crossed grouping factors. Note that the
definition also includes functional random effects $b_{g}(t)$ for a grouping 
variable $g$ with $M$ levels. These are modeled as realizations of a mean-zero 
Gaussian random process on $\{1,\dots,M\} \times \mathcal{T}$ with a general 
covariance function \mbox{$K^{b}(m,m',t,t') = \Cov(b_{g,m}(t), b_{g,m'}(t'))$} 
that is smooth in $t$, where $m, m'$ denote different levels of $g$. Table 
\ref{tab.effects} shows how a selection of the most frequently used effect types
fit into this framework.

We approximate each term $f_r(\mathcal{X}_r, t)$ by a linear combination of basis
functions given by the tensor product of marginal bases for $\mathcal{X}_r$ and
$t$. Since the basis has to be rich enough to ensure sufficient flexibility,
Section \ref{sec:inference} describes a penalized likelihood approach that
stabilizes estimates by suppressing variability of the effects that is not strongly supported by the data and finds a data-driven compromise between goodness of fit and simplicity of the fitted effects.

\begin{table}[!ht]\centering
\caption{A selection of possible model terms $f_r(\mathcal{X}_r, t)$ for model \eqref{eq:fdmodel}. All effects can be constant in $t$ as well.
\label{tab.effects}}
\begin{small}
\begin{tabular}{p{.33\textwidth} p{.3\textwidth} p{.2\textwidth}}
\toprule[0.09 em]
$\mathcal{X}_r$ & type of effect     &    $f_r(\mathcal{X}_r, t)$ \\ 
\hline
$\emptyset$ (none)    & smooth intercept   &                        $\beta_0(t)$ \\
scalar covariate $z$  & linear or smooth  effect    &     $z \beta(t)$, $f(z, t)$ \\
two scalars  $z_1$, $z_2$ & linear or smooth interaction &                       $z_1 z_2 \beta(t)$,  $z_1 f(z_2,t)$, $f(z_1, z_2, t)$\\
\hline 
functional covariate $x(s)$ & linear or smooth (historical) functional effect &     $\int x(s) \beta(s,t)ds$, $\int_{l(t)}^{u(t)}  x(s) \beta(s,t)ds,$ $\int F(x(s), s, t) ds$ \\
functional covariate $v(t)$ & concurrent effects & $v(t)\beta(t)$,  $f(v(t), t)$ \\
functional covariates $v(t), w(t)$ & concurrent interactions & $v(t)w(t)\beta(t)$, $f(v(t), w(t), t)$ \\
\hline
grouping variable $g$ & functional random intercept  &  $b_g(t)$ \\[0.5em]
grouping variable $g$, scalar  $z$ & functional random slope & $z b_{g}(t)$ \\
\hline
curve indicator $i$ & smooth functional residual  &  $e_i(t)$ \\
\bottomrule[0.09 em]
\end{tabular}
\end{small}
\end{table}

\subsection{Data and Notation}
In practice, functional responses $y_i(t)$ are observed on a grid of $T_{i}$ 
points \linebreak \mbox{$\mt_{\bm{i}}=(t_{i1}, \dots,t_{iT_{i}})\tr$} which can be irregular 
and/or sparse. Let $y_{il}=y_i(t_l)$ and \mbox{$\my_{i} = (y_{i1},\dots,y_{i{T_i}}).$}
To fit the model, we form \mbox{$\my = (\my_{1}\tr, \dots, \my_{n}\tr)\tr$} and 
\linebreak\mbox{$\mt = (\mt_{1}\tr, \dots, \mt_{n}\tr)\tr$,} two $N=\sum^n_{i=1} T_i$-vectors that
contain the concatenated observed responses and their argument values, respectively.
Let $\bm{\mathcal{X}}_{\bm{r}i}$ contain the observed values of
$\mathcal{X}_r$ associated with a given $\my_i$. Model \eqref{eq:fdmodel} can
then be expressed as
\begin{equation}
\label{eq:ammodel}
\begin{split}
y_{il} &\sim \mathcal{F}(\mu_{il}, \mnu)\\
g(\mu_{il}) &= \sum^R_{r=1} f_r(\bm{\mathcal{X}}_{\bm{r}i}, t_{il}) 
\end{split}
\end{equation}
for $i= 1, \dots, n$ and $l=1, \dots, T_i$. Let $f(\mt)$ denote the vector of 
function evaluations of  $f$ for each entry in the vector $\mt$ and let $f(\mx, 
\mt)$ denote the vector of evaluations of   $f$ for each combination of rows in 
the vectors or matrices $\mx, \mt$.
In the following, we let $T_i \equiv T$ to simplify notation, but our approach 
is equally suited to data on irregular grids. For regular grids, each 
observed value in any $\mathcal{X}_r$ that is constant over $t$ is simply repeated $T$ times 
to match up with the corresponding entry in the $N=nT$-vector $\my$.

\subsection{Tensor product representation of effects}\label{sec:tensorrep}

We approximate each term $f_r(\mathcal{X}_r, t)$ by a linear combination of basis functions defined on the product space of the two spaces: one for the covariates in $\mathcal{X}_r$ and one over $\mathcal T$, where each marginal basis is
associated with a corresponding marginal penalty. A very versatile method to
construct basis function evaluations on such a joint space is given by the row
tensor product of marginal bases evaluated on $\bm{\mathcal{X}_r}$ and $\mt$
\citep[e.g.][ch.~4.1.8]{Wood2006}. Let $\bm{1}_d=(1,\dots,1)\tr$ denote a 
$d$-vector of ones. The row tensor product of an $m \times a$ matrix $\mA$ and 
an $m \times b$ matrix $\mB$ is defined as the $m\times ab$ matrix $\mA \odot 
\mB = (\mA \otimes \bm{1}_b\tr) \cdot (\bm{1}_a\tr \otimes \mB )$, where $\otimes$ denotes the Kronecker product and $\cdot$ denotes element-wise multiplication.  Specifically, for each of the terms,
\begin{align}
\dimm{f_r(\boldsymbol{\mathcal{X}_r}, \mt)}{N \times 1} &\approx 
\dimm{(\mPhi_{\bm{x}r}}{N
\times K_{xr}} \odot \dimm{\mPhi_{\bm{t}r})}{N \times K_{tr}} 
\dimm{\mtheta_r}{K_{xr} K_{tr} \times 1} = \mPhi_r \mtheta_r,
\label{eq:tensorrep}
\end{align}
where $\mPhi_{\bm{x}r}$ contains the evaluations of a suitable marginal basis for the 
covariate(s) in $\boldsymbol{\mathcal{X}_r}$ and $\mPhi_{\bm{t}r}$ contains the 
evaluations of a marginal basis in $\mt$ with $K_{xr}$ and $K_{tr}$ basis functions, 
respectively. The shape of the function is determined by the vector of  
coefficients $\mtheta_r$. A corresponding penalty term can be defined using the 
Kronecker sum of the marginal penalty matrices $\mP_{\bm{x}r}$ and 
$\mP_{\bm{t}r}$ associated with each basis \citep[ch.~4.1]{Wood2006}, i.e. 
\begin{align}\label{eq:tensorpen}
\begin{split}
\operatorname{pen}(\mtheta_r| \lambda_{tr}, \lambda_{xr}) &=
\mtheta_r^T \mP_{r}(\lambda_{tr},\lambda_{xr}) \mtheta_r, \\
\text{where } \mP_{r}(\lambda_{tr},\lambda_{xr}) &= \lambda_{xr}  \mP_{\bm{x}r} \otimes 
\mI_{K_{tr}}+ \lambda_{tr} \mI_{K_{xr}} \otimes \mP_{\bm{t}r}. 
\end{split}
\end{align}
$\mP_{\bm{x}r}$ and $\mP_{\bm{t}r}$ are known and fixed positive semi-definite penalty matrices 
and $\lambda_{tr}$ and $\lambda_{xr}$ are positive smoothing parameters 
controlling the trade-off between goodness of fit and the smoothness of 
$f_r(\boldsymbol{\mathcal{X}_r}, \mt)$ in $\boldsymbol{\mathcal{X}_r}$ and 
$\mt$, respectively. 

This approach is extremely versatile and powerful as it allows analysts to pick and 
choose bases and penalties best suited to the problem at hand. Any basis and penalty 
over $\mathcal T$ (for example, incorporating monotonicity or periodicity constraints)
can be combined with any basis and penalty over $\mathcal{X}_r$.
We provide some concrete examples for frequently encountered effects:
For a functional intercept $\beta_0(t)$, $\mPhi_{\bm{x}r} = \bm{1}_N$ and $\mP_{\bm{x}r}=0$.
For a linear functional effect of a scalar covariate $z \beta(t)$, $\mPhi_{\bm{x}r} = \bm{z} \otimes \bm 1_T$ and $\mP_{\bm{x}r}=0$, where $\bm{z} = (z_1, \dots, z_n)^T$ contains the observed values of a scalar covariate $z$. 
For a smooth functional effect of a scalar covariate $f(z, t)$, 
$\mPhi_{\bm{x}r}=[\phi_h(z_i)]{\smsubalign{1 &\leq i \leq n \\ 1 &\leq h \leq K_{xr}}} \otimes \bm 1_T$ and $\mP_{\bm{x}r}$ is the penalty associated with the basis functions $\phi_h(z)$.

For a linear effect of a functional covariate $\int_{l(t)}^{u(t)}  x(s) \beta(s,t)ds$,   $$\mPhi_{\bm{x}r}= (\dimm{\bm W_s}{N \times S} \cdot \dimm{\bm X_{\phantom{.}}}{N \times S})  \dimm{\mPhi_s}{S \times K_{xr}},$$ where $\bm W_s$ contains suitable quadrature weights for the numerical integration and zeroes for combinations of $s$ and $t$ not inside the integration range $[l(t), u(t)]$. $\bm X = [x_i(s_k)]{\smsubalign{ 1&\leq i \leq n\\ 1 &\leq k \leq S}} \otimes \bm 1_T$ contains the repeated evaluations of $x(s)$ on the observed grid $(s_1, \dots, s_S)$ and $\mPhi_s = [\phi_h(s_k)]{\smsubalign{1 &\leq k \leq S \\ 1 &\leq h \leq K_{xr}}}$ the evaluations of basis functions $\phi_h(s)$ for the basis expansion of $\beta(s,t)$ over $s$. 
$\mP_{\bm{x}r}$ is again simply the penalty matrix associated with the basis functions $\phi_h(s)$.\\
For a functional random slope $z b_g(t)$ for a grouping factor $g$ with $K_{xr} = M$ levels, 
$\mPhi_{\bm{x}r}= (\operatorname{diag}(\bm z) \mG) \otimes \bm{1}_T$, where 
$\mG = [\delta(g_i = m)]{\smsubalign{1 &\leq i \leq n \\ 1 &\leq m \leq M}}$ is an incidence matrix mapping each observation to its group level. $\mP_{\bm{x}r}$ is typically $\bm I_M$ for $\iid$ random effects, 
but can also be any inverse correlation or covariance matrix defining the expected similarity of levels of $g$ with each other, defined for example via spatial or temporal correlation functions or other similarity measures defined between the levels of $g$.\\
In general, $\mPhi_{\bm{t}r}$ and $\mP_{\bm{t}r}$ can be any matrix of basis function evaluations $\mPhi_{\bm{t}r} = \bm{1}_{n} \otimes  [\phi_k(t_l)]{\smsubalign{1 &\leq l \leq T \\ 1 &\leq k \leq K_{tr}}}$ over $t$ with a corresponding penalty matrix. For effects that are constant over $t$, we simply use $\mPhi_{\bm{t}r} = \bm 1_N$ and $\mP_{\bm{t}r} = 0$.\\   
Note that for data on irregular $t$-grids, the stacking via ``$\otimes \bm{1}_T$'' in the expressions above simply has to be replaced by using a different suitable repetition pattern. \citet{famm2014} discuss tensor product representations of many additional effects available in our implementation. 

As for other penalized spline models, fits are usually robust against increases in $K_{tr}$ and $K_{xr}$ once a sufficiently large number of basis functions has been chosen since it is the penalty parameters $\lambda_{tr}, \lambda_{xr}$ that control the
effective degrees of freedom of the fit, c.f.~\citet{Ruppert2002}, \citet[Ch. 4.1.7]{Wood2006}. \citet{Pya2016} describe and implement a test procedure to determine sufficiently large basis dimensions that is also applicable to our model class.

\subsection{Inference and Implementation}\label{sec:inference}

Let $\mPhi = [\mPhi_1 | \dots | \mPhi_R]$ contain the  concatenated design matrices
$\mPhi_r$ associated with the different model terms and $\mtheta=(\mtheta_1\tr, \dots 
,\mtheta_R\tr)\tr$ the respective stacked coefficient vectors $\mtheta_r$. Let 
$\mlambda = (\lambda_{t1},\lambda_{x1}, \dots, \lambda_{tR},\lambda_{xR})$. 
The penalized log-likelihood to be maximized is given by
\begin{align}
\ell_p(\mtheta, \mlambda, \mnu | \my) &= \ell(\bm{\mu}, \mnu|\my) - \tfrac{1}{2} 
\sum^R_{r=1} \operatorname{pen}(\mtheta_r| \lambda_{tr}, \lambda_{xr})
\label{eq:jointlik}
\end{align}
where $\bm\mu =  g^{-1}(\mPhi \mtheta)$ and $\ell(\bm{\mu}, \mnu|\my) = \sum^n_{i=1}\sum^{T_i}_{l=1} \ell(\mu_{il}, \mnu | y_{il})$ is the log-likelihood function implied by the respective response distribution 
$\mathcal{F}(\mu_{il},\bm\nu)$ given in \eqref{eq:ammodel}.

Recent results indicate  that REML-based inference for this class of penalized regression models can be preferable to GCV-based optimization \citep[][Section 4]{ReissOgden2009, Wood2011}. \citet{Wood2014} describes a numerically stable implementation for directly optimizing a Laplace-approximate marginal likelihood of the smoothing parameters $\mlambda$ which we summarize below. Note that this is equivalent to REML optimization in the Gaussian response case. The idea is to iteratively perform optimization over $\mlambda$ and then estimate basis coefficients $\mtheta$ given $\mlambda$ via standard penalized likelihood methods.\\

We first define a Laplace approximation $\ell_{LA}(\mlambda,\mnu | \my) \approx \int \ell_p(\mtheta,\mlambda,\mnu| \my) d\mtheta$ to the marginal (restricted) ML score suitable for optimization. \newline Let \mbox{$\mH = - \partial^2 \ell_p(\mtheta, \mlambda, \mnu | \my) /  (\partial \mtheta \partial \mtheta^T)$} denote the negative Hessian of the penalized log-likelihood \eqref{eq:jointlik}, and let $\tilde\mtheta = \arg\max_\theta \ell_p(\mtheta,\mlambda,\mnu | \my)$ for fixed $\mlambda, \mnu$ with $\tilde{\bm{\mu}} = g^{-1}(\mPhi \tilde\mtheta)$. The approximate marginal (restricted) ML criterion can then be defined as 
\begin{align}
\begin{split}
\ell_{LA}(\mlambda,\mnu | \my) = \ell(\tilde{\bm{\mu}}, \mnu | \my) -  \tfrac{1}{2} 
\sum^R_{r=1}  \left(\operatorname{pen}(\tilde\mtheta_r| \lambda_{tr}, 
\lambda_{xr})   -  \log|\mP_r(\lambda_{tr}, \lambda_{xr})|^+\right) -\\
  \quad\tfrac{1}{2}\log\left|\mH\right| + \tfrac{1}{2} d_\emptyset \log(2\pi), 
\end{split}
\label{eq:laplacereml}
\end{align}
where $|\mA|^+$ denotes the generalized determinant, i.e., the product of the positive eigenvalues of $\mA$, and $d_\emptyset$ is the sum of the rank deficiencies of the $\mP_r(\lambda_{tr}, \lambda_{xr}), r=1, \dots, R$.

The R package \code{mgcv} \citep{Wood2006} provides an efficient and numerically stable implementation for the iterative optimization of \eqref{eq:laplacereml} that 1) finds $\tilde\mtheta$ for each candidate $\mlambda$ via penalized iteratively re-weighted least squares, 2) computes the gradient and Hessian of $\ell_{LA}(\mlambda,\mnu | \my)$ w.r.t.~$\log(\mlambda)$ via implicit differentiation, 3) updates $\log(\mlambda)$ via Newton's method and 4) estimates $\mnu$ either as part of the maximization for $\mlambda$ or based on the Pearson statistic of the fitted model. See \citet{Wood2011, Wood2014, WoodPyaSaefken2015} for technical details and numerical considerations.

Confidence intervals \citep{MarraWood2012} and tests \citep{Wood2013} for the estimated effects are based on the asymptotic normality of ML estimates. Confidence intervals use the variance of the distance between the estimators and true effects. Similar to the idea behind the MSE, this accounts for both the variance and the penalization induced bias (shrinkage effect) of the spline estimators. 

We provide an implementation of the full generality of the proposed models in the \code{pffr()} function in the R package \code{refund} \citep{refund} and use the \code{mgcv} \citep{Wood2011} implementation of (scalar) generalized additive mixed models as computational backend. The \code{pffr()} function provides a formula-based wrapper for the definition of a broad variety of functional effects as in Section \ref{sec:tensorrep} as well as convenience functions for summarizing and visualizing model fits and generating predictions.

%% file: gfamm_data.tex
\section{Application to Pigs' Feeding Behavior}\label{sec:app}
\subsection{Data}\label{sec:app-data}

We use data on pigs' feeding behavior collected in the ICT-AGRI era-net project
``PIGWISE'', funded by the European Union. In this project, high frequency radio frequency identification (HF RFID) antennas were
installed above the troughs of pig barns to register the feeding times of pigs
equipped with passive RFID tags \citep{Maselyne2014}.
We are interested in modelling the binary functional data that arises from measuring
the proximity of the pig to the trough (yes-no) every 10 secs over 102 days available
for each pig. The raw data for one such pig, called `pig 57' whose behavior we analyse in depth is shown in Figure~\ref{fig:piggyplot1} in Appendix \ref{app:alternative-models}.  
Such models of (a proxy of) individual feeding behavior can be useful for ethology research as well as monitoring individual pigs' health status and/or quality of the available feed stock, c.f. \citet{Gertheiss2014}. Available covariates include the time of day, the age of the pig (i.e, the day of observation), as well as the barn's temperature and humidity.

\subsection{Model}\label{sec:app-model}

Due to pronounced differences in feeding behavior between individual pigs \citep[c.f. ][Figure 4]{Gertheiss2014}, we focus on pig-wise models and model the observed feeding rate for a single pig for each day $i=1,\dots,102$ as a smooth function over the daytime $t$.
For our model, we aggregate the originally observed binary feeding indicators
$\tilde y_i(t)$, which are measured every 10 seconds, over 10 min intervals into
$y_i(t)$. This temporal resolution is sufficient for the purposes of our modeling
effort. We then model
\begin{align*} 
y_i(t) |\X_i &\sim \text{Bin}(60, \pi=\mu_i(t));\\
\mu_i(t) &= \text{logit}^{-1}\left(\sum_{r=1}^R f_r(\X_{ri},t)\right). 
\end{align*}
As is typical for applied problems such as this, many different distributional
assumptions for $y_i(t) |\X_i$ and specifications of  $\sum_{r=1}^R f_r(\X_{ri},t)$
are conceivable. In this case, both Beta and Negative Binomial distributions
yielded less plausible and more unstable fits than the Binomial distribution. With
respect to the additive predictor, the effect of temperature or humidity (i.e., 
$\X_{ri} = \texttt{hum}_i(t)$), for example, could be modeled as a linear functional
effect $\int^{u(t)}_{l(t)} \texttt{hum}(s) \beta(s,t) ds$, which represents the cumulative
effect of humidity exposure over the time window $l(t) \leq t \leq  u(t)$, or the concurrent
effect of humidity, either linearly via $\texttt{hum}(t)\beta_h(t)$ or non-linearly via
$f(\texttt{hum}(t), t)$. Auto-regressive and lagged effects of previous feeding behavior 
(i.e., $\X_{ri} = y_i(t)$) could be modeled similarly either as the cumulative effect
of previous feeding over a certain time window ($\int^{t-10\text{min}}_{t-\delta} y(s) \beta(s,t) ds$)
or nonlinear ($f(y_i(t-\delta), t)$) or linear ($y_i(t-\delta)\beta(t)$)
effects of the lagged responses with a pre-specified time lag $\delta$. Finally,
our implementation also offers diverse possibilities for accounting for aging effects
and day-to-day variations in behavior (i.e., $\X_{ri} = i$). Aging effects that result
in a gradual change of daily feeding rates over time could be represented as 
$i \beta(t)$ for a gradual linear change or as $f(i, t)$ for a nonlinear smooth change over days $i$. Less systematic day-to-day variations can be modeled as daily
functional random intercepts $b_i(t)$, potentially auto-correlated over $i$ to
encourage similar shapes for neighboring days. In this application, the performances of most of these models on the validation set were fairly similar and the subsequent section presents results for a methodologically interesting model that was among the most accurate models on the validation set.

\subsection{Results}\label{sec:app-res}

In what follows, we present detailed results for an auto-regressive binomial logit
model with smoothly varying day effects for pig 57 (see Figure~\ref{fig:piggyplot1}, Appendix \ref{app:alternative-models}):
$$\text{logit}\left(\mu_i(t)\right) = \beta_0(t) + f(i, t) + \int^{t-10\text{min}}_{t-3\text{h}} y_i(s) \beta(t,s)ds.$$
This model assumes that feeding behavior during the previous 3 hours affects the
current feeding rate. We use periodic P-spline bases over $t$ to enforce similar $f_r(\X_{ri}, t)$
for $t=$00:00h and  $t=$23:59h for all $r.$ The term $f(i,t)$ can be interpreted
as a non-linear aging effect on the feeding rates or as a functional random day effect
whose shapes vary smoothly across days $i$.  This model contains $\operatorname{dim}(\mlambda) = 5$
smoothing parameters (1 for the functional intercept, 2 each for the daily and auto-regressive effects),
$\operatorname{dim}(\mtheta) = 106$ spline coefficients (24 for the intercept,
25 for the autoregressive effect, 56 for the day effect) and $N = nT = 9648$ observations
in the training data. We leave out every third day to serve as external validation data for evaluating the generalizability of the model. The fit takes about 1 minute on a modern desktop PC. Appendix~\ref{app:alternative-models} contains detailed results for an alternative model specification, Section~\ref{sec:sim-binomial} describes results for synthetic data with similar structure and size.

Beside giving insight into pigs' feeding patterns and how these patterns change as the pigs age, the model reported here enables short-term predictions of feeding probabilities for a given pig based on its previous feeding behavior. This could be very helpful when using the RFID system for surveillance and early identification of pigs showing unusual feeding behavior. Unusual feeding behavior is then indicated by model predictions that are consistently wrong; i.e, the pig is not behaving as expected. 
Such discrepancies can then indicate problems such as disease or low-quality feed stock. For the auto-regressive model discussed here, only very short-term predictions 10 minutes into the future can be generated easily as $y_i(t-10\text{min})$ is required as a covariate value. Other model formulations without such auto-regressive terms or larger lead times of auto-regressive terms will allow more long-term forecasting. More long-term forecasts for auto-regressive models could also be achieved by using predictions $\hat y_i(t)$ instead of observed values of $y_i(t)$ as inputs for the forecast. These can be generated sequentially for timepoints farther and farther in the future, but errors and uncertainties will accumulate correspondingly. A detailed analysis of this issue is beyond the scope of this paper, \citet{Tashman2000} gives an overview. In addition to the lagged response value, also the aging effect $f(i,t)$ is needed. Here, we could either use the value from the preceding day $i-1$, if it can be assumed that the pig's (expected) behavior does not change substantially from one day to the next; or do some extrapolation.

\begin{knitrout}\scriptsize
\definecolor{shadecolor}{rgb}{0.969, 0.969, 0.969}\color{fgcolor}\begin{figure}

{\centering \includegraphics[width=\textwidth]{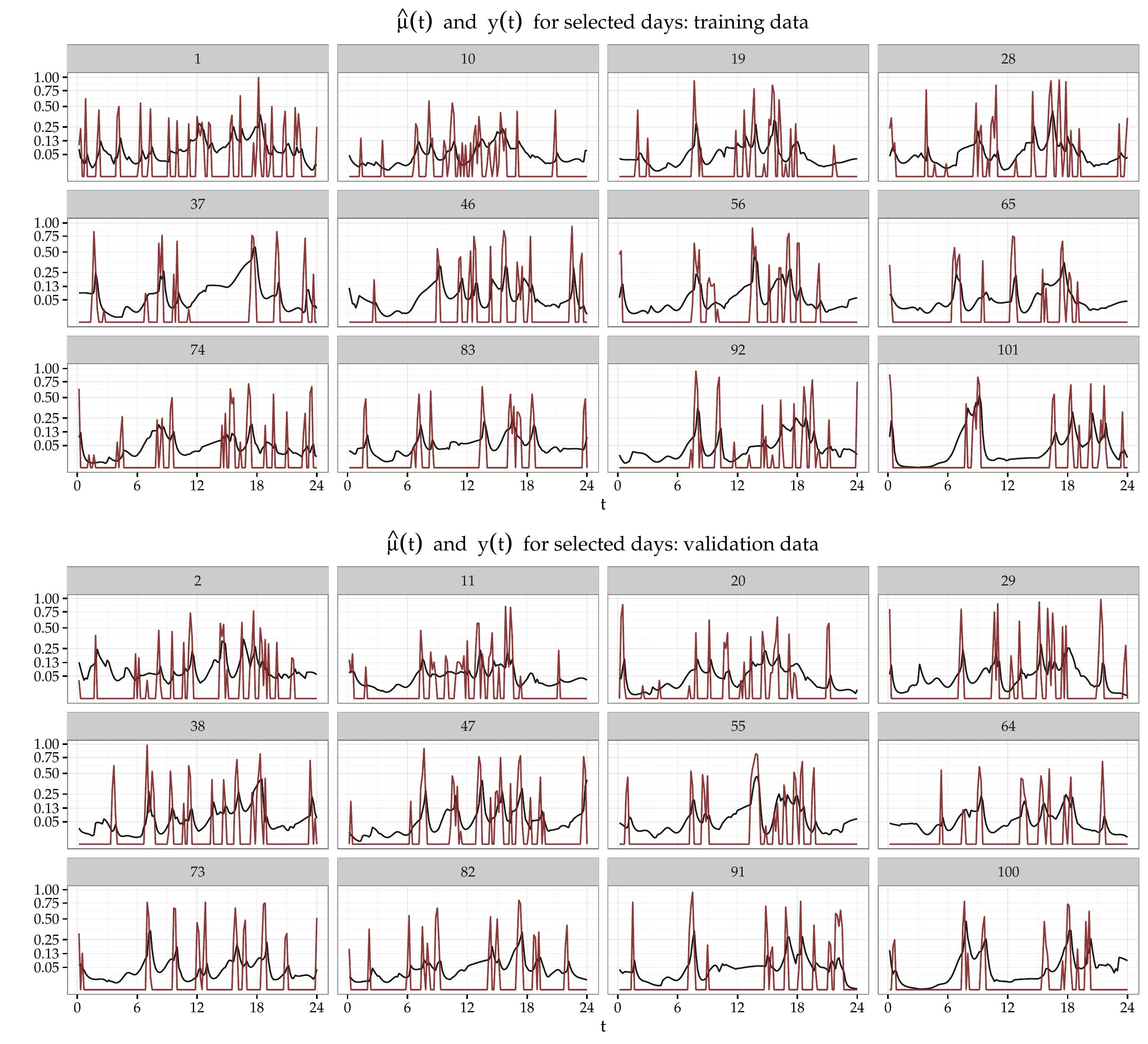} 

}

\caption{Training (top) and validation data (bottom) results for pig 57 for selected days with a smoothly varying day effect. Black lines for fitted (top) and predicted (bottom) values $\hat\mu_i(t)$, red for observed feeding rate $y_i(t)/60$. Numbers above each panel denote the day. Vertical axis on $\sqrt{\;}$-scale.}\label{fig:plot_m64_fit}
\end{figure}

\end{knitrout}
Figure~\ref{fig:plot_m64_fit} shows fitted and observed values for 12 selected days
from the training data (top) as well as estimated and observed values for 12 selected
days from the validation data (bottom). The model is able to reproduce many of the
feeding episodes, in the sense that peaks in the estimated probability curves mostly
line up well with observed spikes of $y_i(t)$. This model explains about 24\% of the
deviance and, taking the mean over all days and time-points, achieves a Brier score of
about 0.024 on both the training data and the validation data,
i.e., we see no evidence for overfitting with this model specification.
This model performs somewhat better on the validation data than an alternative model
specification with $\iid$ random functional day effects $b_i(t)$, c.f.~Appendix~\ref{app:alternative-models}, Figure~\ref{fig:plot_m62_fit}.

\begin{knitrout}\scriptsize
\definecolor{shadecolor}{rgb}{0.969, 0.969, 0.969}\color{fgcolor}\begin{figure}

{\centering \includegraphics[width=\textwidth]{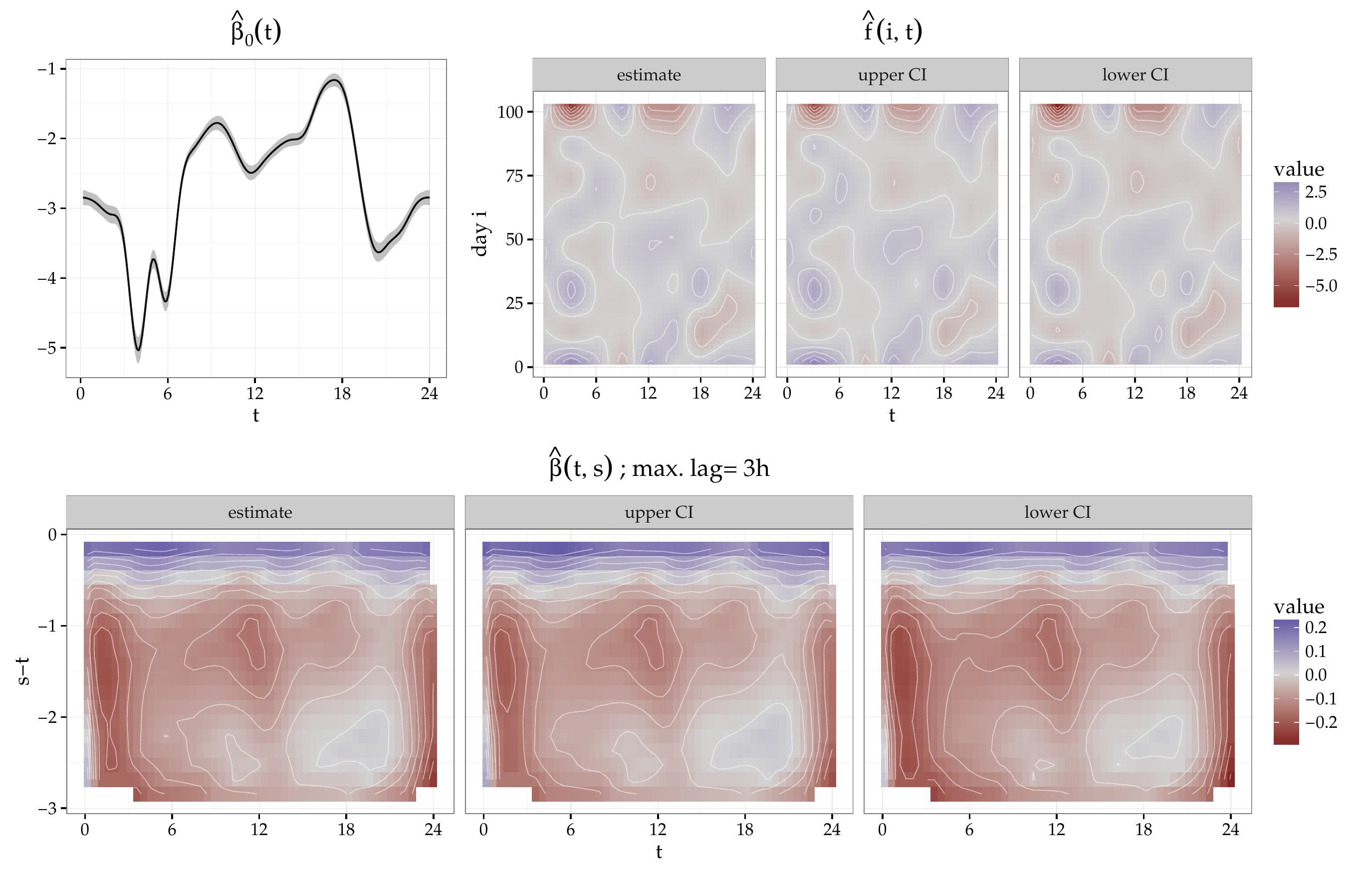} 

}

\caption[Top row]{Top row: Estimated functional intercept and smoothly varying day effects for the model with smoothly varying day effect.\\ Bottom: Estimated coefficient surface for cumulative auto-regressive effect. Point-wise intervals are  $\pm$ 2 standard errors.}\label{fig:plot_m64_coef}
\end{figure}

\end{knitrout}

Figure~\ref{fig:plot_m64_coef} shows the estimated components of the additive predictor for the model with smoothly varying day effects.
The functional intercept in the left top panel of Figure~\ref{fig:plot_m64_coef} shows two clear peaks of the feeding rate around 10h and 18h, and very low feeding activity from 19h to 23h and from 2h to 6h. A two-peak profile has been identified previously as the typical feeding behavior of pigs \citep{Montgomery1978, Guillemet2006}.
Our analysis exploiting the rich information provided by the new RFID antennas used \citep{Maselyne2014}, however, gives pig-specific results, whereas previous studies typically reported on group characteristics; see, e.g., \citet{Hyun1997} and \citet{Guillemet2006}. Furthermore, existing results usually aggregated over time periods.

The estimated smoothly varying day effect (top right panel) shows increased feeding
rates in the early morning in the first few days and a corresponding strong reduction in feeding rates in the early morning towards the end of the fattening period, as well as a tendency for increased feeding activity to concentrate more strongly in two periods around 9h and 21h towards the end of the observation period, a pattern also visible in the raw data shown in Figure~\ref{fig:piggyplot1} (Appendix \ref{app:alternative-models}). The cumulative auto-regressive
effect of feeding during the previous 3 hours is displayed in the bottom row of Figure~\ref{fig:plot_m64_coef}. 
Unsurprisingly, feeding behavior in the immediate past is associated positively
with current feeding (c.f.~the blue region for $s-t \approx 0$), especially during off-peak times in the early morning where a higher propensity for feeding is not already modeled by the global functional intercept.
The model also finds a negative association between prior feeding during the night and
feeding in the very early hours of the morning that takes place 1 to 3 hours later (c.f.~the red region around $t=0$). Confidence intervals for all three effects show that  they can be estimated fairly precisely.

%% file: gfamm_sim.tex
\section{Simulation Study}\label{sec:sim}

This section describes results on binomial
(Section \ref{sec:sim-binomial}) and Beta-, $t$- and negative binomial-distributed   data (\ref{sec:sim-families}). Sections \ref{app:wang} and \ref{app:goldsmith} present replications of (parts of) the simulation studies on functional random intercept models in \citet{WangShi2014} and \citet{Goldsmith2015}, respectively. Note that neither of these competing approaches are able to fit nonlinear effects of scalar covariates, effects of functional covariates, functional random slopes or correlated functional random effects in their current implementation, while \code{pffr} does. Both approaches are implemented only for binary data, not the broad class of distributions available for \code{pffr}-models.
All results presented in this section can be reproduced with the code supplement available for this work at \verb+http://stat.uni-muenchen.de/~scheipl/downloads/gfamm-code.zip+.

To evaluate results on artificial data, we use the relative root integrated mean square error rRIMSE of the estimates evaluated on equidistant grid points over $\mathcal T$:
$$
\text{rRIMSE}(\hat f_r(\mathcal{X}_r, t)) = \sqrt{\tfrac{|\mathcal T|}{nT}\sum_{i,l}\frac{(\hat f_r(\bm{\mathcal{X}}_{\bm{r}i}, t_l) - f(\bm{\mathcal{X}}_{\bm{r}i}, t_l))^2}{\text{sd}(f(\bm{\mathcal{X}}_{\bm{r}i}, \bm t))^2}},
$$
where $\text{sd}(f(\bm{\mathcal{X}}_{\bm{r}i}, \bm t))$ is the empirical standard deviation of $[f(\bm{\mathcal{X}}_{\bm{r}i}, t_l)]_{l=1, \dots, T}$, i.e., the variability of the estimand values over $\mt$. We use the relative RIMSE rather than RIMSE$(\hat f_r(\mathcal{X}_r, t)) =   \sqrt{\tfrac{|\mathcal T|}{nT}\sum_{i,l} (\hat f_r(\bm{\mathcal{X}}_{\bm{r}i}, t_l) - f_r(\bm{\mathcal{X}}_{\bm{r}i}, t_l))^2}$ in order to make results more comparable between estimands with different numerical ranges and to give a more intuitively accessible measure of the estimation error, i.e., the relative size of the estimation error compared to the variability of the estimand across $\mathcal T$.


\input{gfamm_sim_binomial.tex}


\input{gfamm_sim_families.tex}


\input{appendix_wang.tex}


\input{appendix_goldsmith.tex}

\subsection{Summary of Simulation results}
We achieve useful and mostly reliable estimates for many complex settings in data structured like the PIGWISE data in acceptable time. Simulation results for less challenging settings and other response distributions than Binomial are good to excellent as well.
Our partial replications of the simulation studies in \citet{WangShi2014} and \citet{Goldsmith2015} in Sections \ref{app:wang} and \ref{app:goldsmith}, respectively, show that
our implementation achieves highly competitive results in similar or much shorter computation times even though the settings are tailored towards the competing earlier approaches which are also less flexible than ours by quite a large margin.

%% file: gfamm_sim_binomial.tex
\subsection{Synthetic Data: Binomial Additive Mixed Model}\label{sec:sim-binomial}
We describe an extensive simulation study to evaluate the quality of estimation for 
various potential model specifications for the PIGWISE data presented in Section \ref{sec:app-data}. We simulate Binomial data $y_i(t) \sim B\left(\text{trials}, \pi=\logit^{-1}\left(\eta_i(t)\right)\right)$ with $\text{trials}=10, 60, 200$ for $n=100$ observations on $T=150$ equidistant grid-points over $[0, 1]$ for a variety of true additive predictors $\eta_i(t) =  \sum_{r=1}^R f_r(\X_{ri},t)$. The subsequent paragraph describes the various $f_r(\X_{ri},t)$ we investigate.

The additive predictor always includes a functional intercept $\beta_0(t)$. 
For modelling day-to-day differences in feeding rates, the additive predictor includes an $\iid$ functional random intercept $b_i(t)$ (setting: \code{ri}) or a smoothly varying day effect $f(i, t)$ (setting: \code{day}).
Note that the ``true'' functional random intercepts were generated with Laplace-distributed spline coefficients in order to mimic the spiky nature of the application data. For modeling the auto-regressive effect of past feeding behavior, the additive predictor can include one of
\begin{compactitem}
\item a time-varying nonlinear term $f(\tilde y_i(t-.005), t)$ (\code{lag}),
\item a time-constant nonlinear term $f(\tilde y_i(t-.005))$ (\code{lag.c}),
\item a cumulative linear term over the previous 0.3 time units 
$\int_W \tilde y_i(s) \beta(t, s) ds$; $W= \{s: t-0.3 \leq s < t\}$ (\code{ff.3}),
\item a cumulative linear term over the previous 0.6 time units
$\int_W \tilde y_i(s) \beta(t, s) ds$; $W= \{s: t-0.6 \leq s < t\}$ (\code{ff.6}),
\end{compactitem}
where $\tilde y_i(s) = (y_i(s) - \bar y(s))$ denotes the centered responses and $\bar y(s)$
the mean response function.
The simulated data also contains humidity and temperature profiles generated 
by drawing random FPC score vectors from the respective empirical FPC score distributions estimated from the (centered) humidity and temperature data provided in the PIGWISE data. For modeling the effect of these functional covariates, the additive predictor includes either a nonlinear concurrent interaction effect 
$f(\text{hum}_i(t), \text{temp}_i(t), t)$ (\code{ht}) or a nonlinear time-constant concurrent interaction effect $f(\text{hum}_i(t), \text{temp}_i(t))$  (\code{ht.c}).

We simulate data based on additive predictors containing each of the terms described above on its own, as well as additive predictors containing either 
functional random intercepts (\code{ri}) or a smoothly varying day effect (\code{day}) and one of the remaining terms, for a total of 21 different combinations. 
For all of these settings, we vary the amplitude of the global functional intercept (small: $\logit^{-1}(\beta_0(t)) \in [0.06, 0.13]\;\forall\, t$; intermediate: $\in [0.04, 0.19]$; large: $\in [0.02, 0.34]$). For the settings with just a single additional term we also varied the absolute effect size of these terms. 
For small effect sizes the respective 
$\exp(f_r(\X_{ri},t)) \in [0.5, 2]\;\forall\, t$, is $\in [0.2, 5]$ for intermediate effect sizes and $\in [0.1, 10]$ for large effect sizes. We ran 30 replicates for each of the resulting 333 combinations of simulation parameters, for a total of 9990 fitted models. Since the settings' effects are large and systematic and we use a fully crossed experimental design, 30 replicates are more than enough to draw reliable qualitative conclusions from the results here. 

We focus on the results for $\text{trials}=60$ with intermediate
amplitude of the global intercept as these settings are most similar to the application. Results for the other settings are qualitatively similar, with the expected trends: performance improves for more informative data (i.e, more ``trials'') and larger effect sizes and worsens for less informative data with  fewer ``trials'' and smaller effect sizes.

\begin{knitrout}\scriptsize
\definecolor{shadecolor}{rgb}{0.969, 0.969, 0.969}\color{fgcolor}\begin{figure}

{\centering \includegraphics[width=\textwidth]{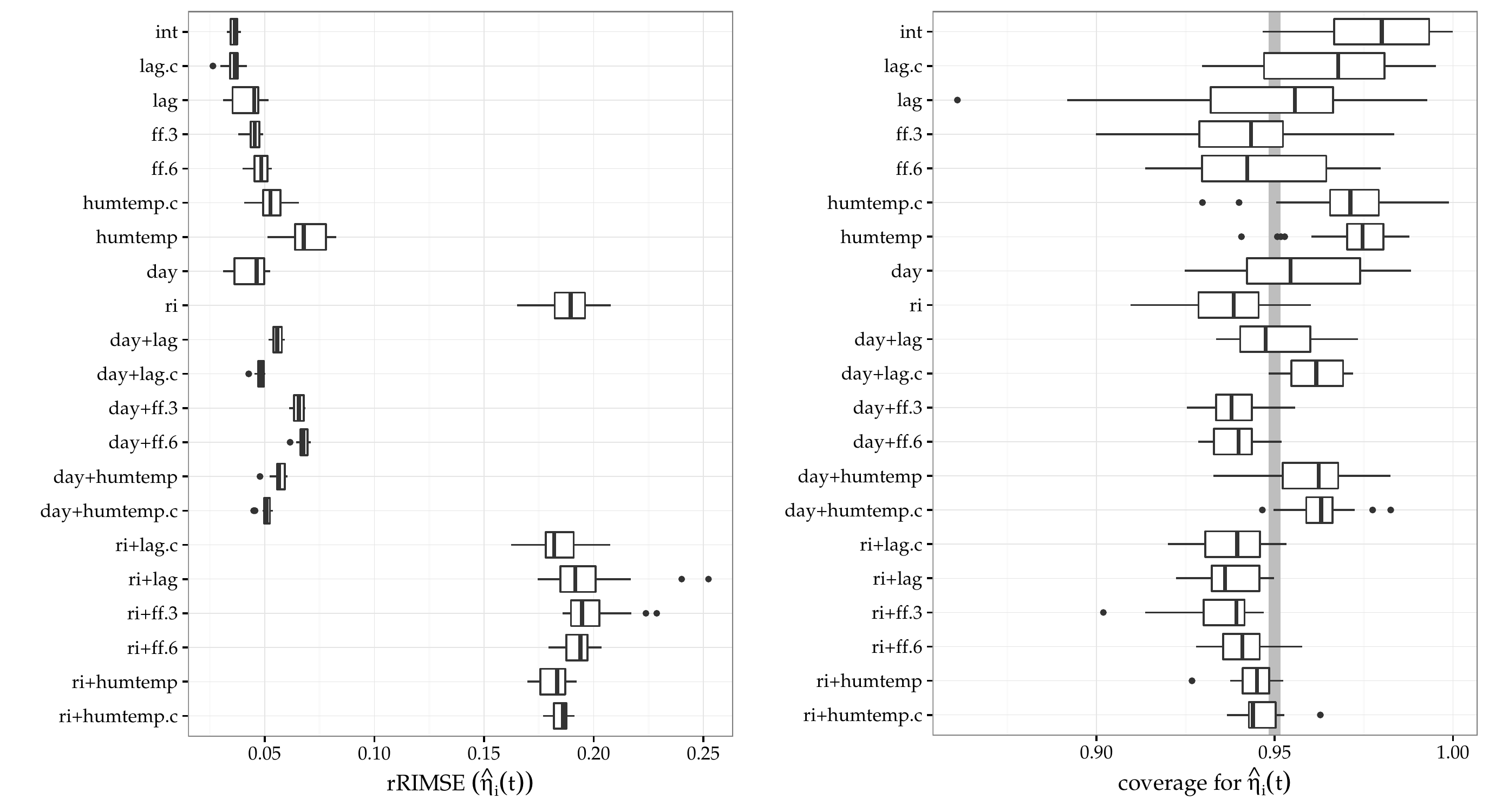} 

}

\caption[rRIMSE]{rRIMSE$(\hat\eta(t))$ and coverages for the different models for synthetic binomial data (60 trials, $\logit^{-1}(\beta_0(t)) \in [0.04, 0.19]$).}\label{fig:binomial-eta-rmse-cover}
\end{figure}

\end{knitrout}
Figure \ref{fig:binomial-eta-rmse-cover} shows rRIMSE$(\hat\eta(t))$ (left panel) and average point-wise coverages (right panel) for the various settings  -- it is easy to see that the concurrent functional effects (\code{humtemp, humtemp.c})
and especially the daily random intercepts (\code{ri}) are the hardest to estimate.
Further study shows that concurrent interaction effects are very sensitive to the rank of the functional covariates involved: both humidity and temperature are practically constant in this data and show little variability, so effects are hard to estimate reliably in this special case due to a lack of variability in the covariate. The daily random intercepts are generated from a Laplace distribution of the coefficients to yield ``peaky'' functions as seen in the application, not the Gaussian distribution our model assumes. Figure \ref{fig:binomial-riff6-ex} in Appendix \ref{app:ex_binomial} displays a typical fit for the setting with functional random intercepts and a cumulative autoregressive effect (\code{ri+ff.6}) and shows that rRIMSEs around .2 correspond to useful and sensible fits in this setting. Also note that performance improves dramatically if the Gaussianity assumption for the random intercept functions is met, c.f.~the results in Sections \ref{app:wang} and \ref{app:goldsmith} for Gaussian random effects in some much simpler settings.  

Note that we are trying to estimate a latent smooth function for each day based only on the feeding episodes for that day for \code{ri} and, through smoothing, based on each day and its neighboring days in the case of \code{day}. The results show that this strategy can work rather well, albeit with a high computational cost in the case of \code{ri}: On our hardware, models with such functional random intercepts usually take between 5 minutes and 2 hours to fit, while models with a smooth day effect usually take between 15 seconds and 15 minutes, depending on which covariate effects are present in the model. See Figure \ref{fig:binomial-time} in Appendix \ref{app:ex_binomial} for details on computation times.

As seen in the replication studies in Sections \ref{app:wang} and \ref{app:goldsmith}, estimating random intercepts with our approach also works as well or better than competing approaches in more conventional settings with multiple replicates per group level rather than the one considered here with a single curve per group level. The results here still represent quite successful fits -- Figure \ref{fig:binomial-dayff6-ex} shows results for a typical \code{day+ff.6}-model that yields about the median RIMSE for that setting. Most estimates are very close to the ``truth''.

In terms of CI coverage, we find that CIs for $\hat\eta_i(t)$ (Figure \ref{fig:binomial-eta-rmse-cover}, right panel) are usually close to their nominal $95$\% level.

\begin{knitrout}\scriptsize
\definecolor{shadecolor}{rgb}{0.969, 0.969, 0.969}\color{fgcolor}\begin{figure}

{\centering \includegraphics[width=\textwidth]{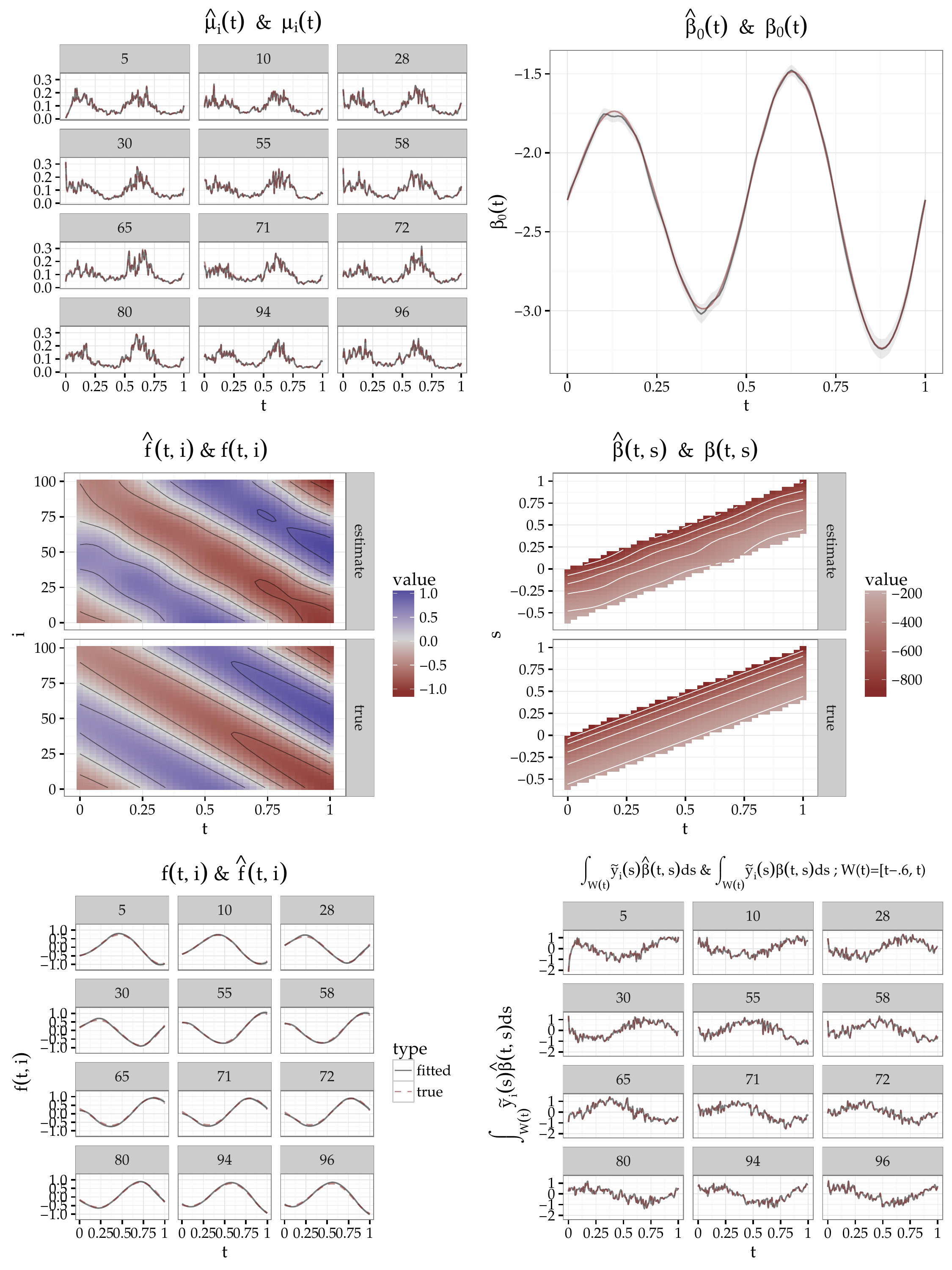} 

}

\caption{Typical model fit for setting \code{day+ff.6}: smoothly varying day effect with a cumulative auto-regressive term. Left column shows true (red, dashed) and estimated (black, solid) conditional expectation (top) for 12 randomly selected days in the top row, estimated (top panel) and true (bottom panel) smoothly varying day-daytime effect (middle and bottom rows). Right column shows true and estimated global intercept (top), estimated (middle, top panel) and true (middle, bottom panel) coefficient surface for the cumulative auto-regressive effect and true and estimated contributions of the cumulative auto-regressive effect to the additive predictor (bottom). For this model, rRIMSE$(\hat\eta(t)) =  0.054 $, while the median rRIMSE for this setting is 0.069}\label{fig:binomial-dayff6-ex}
\end{figure}

\end{knitrout}
Appendix \ref{app:ex_binomial} shows typical fits for some of the remaining settings. In summary, the proposed approach yields useful point estimates for most terms under consideration and confidence intervals with coverages consistently close to the nominal level on data with a similar structure as the application.

%% file: gfamm_sim_families.tex
\subsection{Synthetic Data: Beta-, Negative Binomial-, $t(3)$-Distribution}\label{sec:sim-families}

\subsubsection{Data Generating Process} 
We generate data with responses with scaled $t(3)$-distributed errors,
Beta($\alpha, \beta$)-distributed responses and Negative Binomial NB($\mu, \theta$)-distributed responses. We generate data with $n=100, 300$ observations, each with signal-to-noise ratios SNR $= 1, 5$. SNR is not easily adjustable for Negative Binomial data since variance and mean cannot be specified separately. In our simulation, we use $g(\cdot)=\log(\cdot)$, with $\theta= 0.5$, so that $\Var(y(t)) = \EV(y(t)) + 2 \EV(y(t))^2$. See Appendix \ref{app:ex-families} for details.

Functional responses are evaluated on $T=60$ equidistant grid-points over $[0, 1]$. We investigate performance for 4 different additive predictors:
\begin{compactitem}
\item \code{int}: $\EV(y(t)) = g^{-1}(\beta_0(t))$
\item \code{smoo}: index-varying non-linear effect: $\EV(y(t)) = g^{-1}(\beta_0(t) + f(x, t))$ 
\item \code{te}: non-linear interaction: $\EV(y(t)) = g^{-1}(\beta_0(t) + f(x_1, x_2))$ via tensor product spline
\item \code{ff}:  functional covariate: $\EV(y(t)) = g^{-1}(\beta_0(t) + \int x(s)\beta(t, s) ds)$
\end{compactitem}
50 replicates were run for each of the 32 settings for Beta and $t(3)$ and each of the 16 settings for NB.

\subsubsection{Results} 

\begin{knitrout}\scriptsize
\definecolor{shadecolor}{rgb}{0.969, 0.969, 0.969}\color{fgcolor}\begin{figure}

{\centering \includegraphics[width=\textwidth]{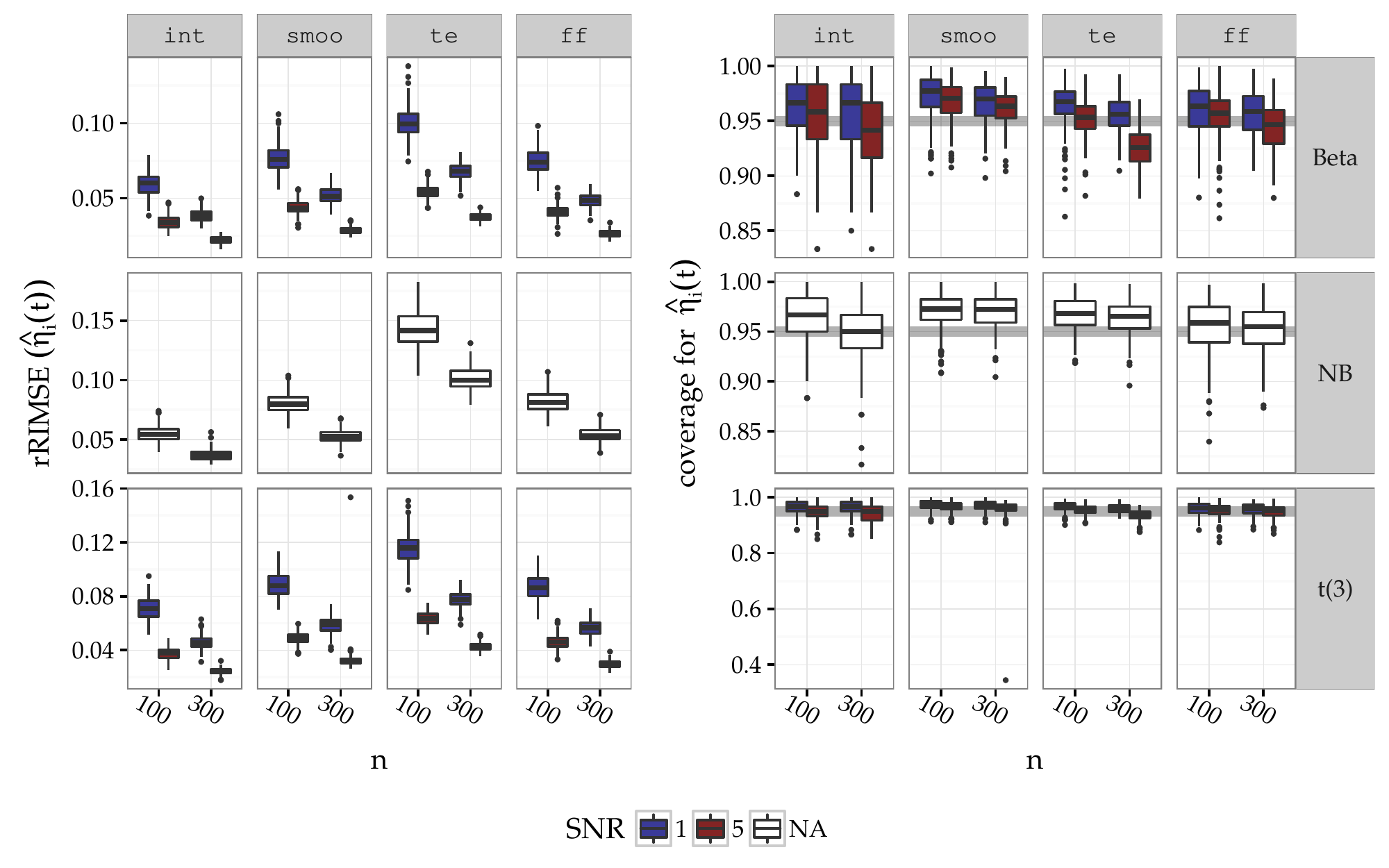} 

}

\caption[Relative RIMSE (left) and achieved coverage (right, nominal 95\%) for estimated additive predictor]{Relative RIMSE (left) and achieved coverage (right, nominal 95\%) for estimated additive predictor. Rows for the three distributions and columns for the 4 model settings. Blue for high noise settings with SNR $= 1$, red for SNR $= 5$, white for NB without available SNR. Horizontal axis represents the varying number of observations. Fat grey horizonal line denotes nominal 95\% coverage.}\label{fig:families-eta}
\end{figure}

\end{knitrout}
\begin{knitrout}\scriptsize
\definecolor{shadecolor}{rgb}{0.969, 0.969, 0.969}\color{fgcolor}\begin{figure}

{\centering \includegraphics[width=\textwidth]{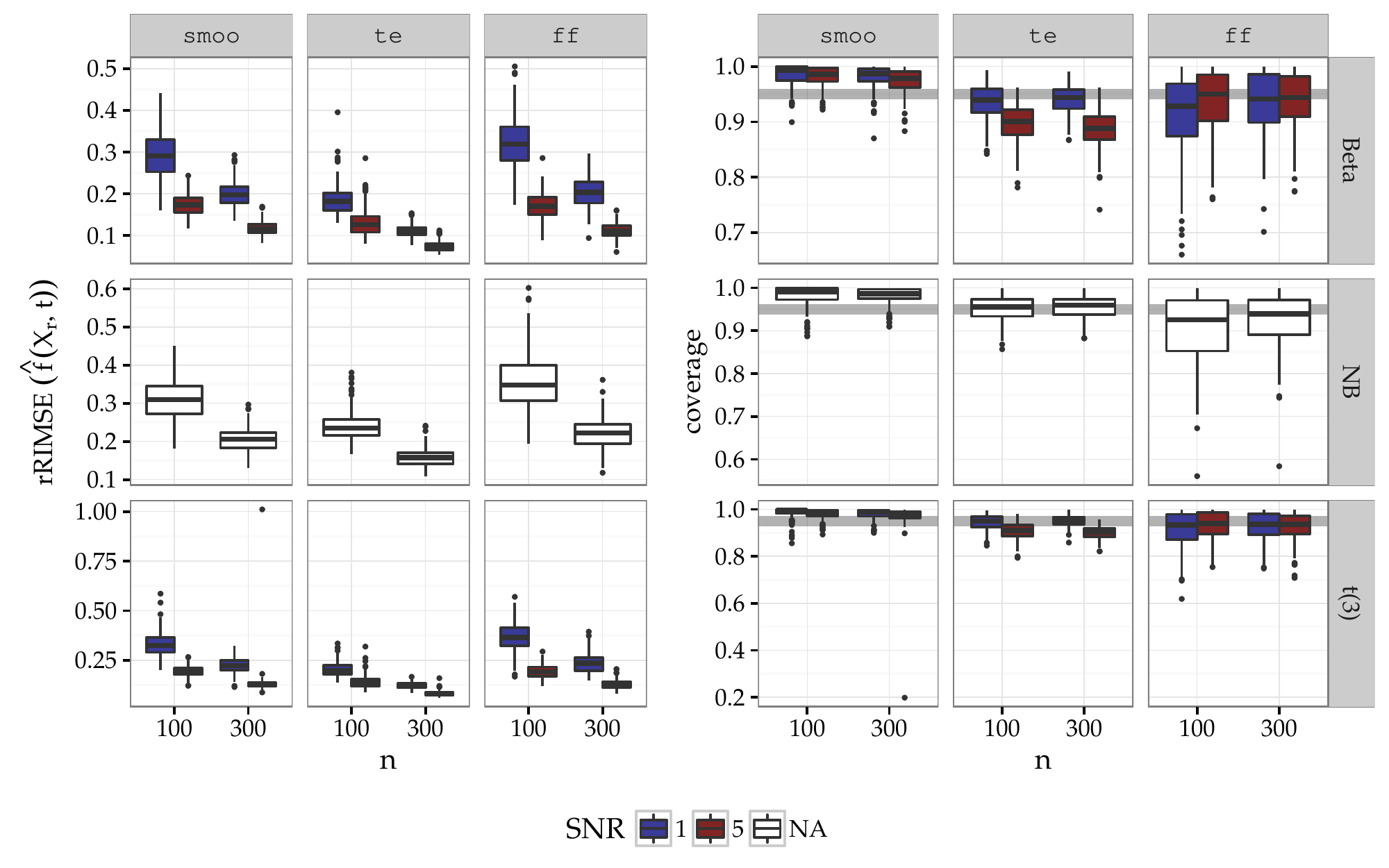} 

}

\caption[Relative RIMSE (left) and achieved coverage (right, nominal 95\%) for estimated effects]{Relative RIMSE (left) and achieved coverage (right, nominal 95\%) for estimated effects. Columns for the three distributions and rows for settings with index-varying nonlinear effect, nonlinear interaction effect and linear function-on-function effect, respectively. Blue for high noise settings with SNR $= 1$, red for SNR $= 5$, white for NB without available SNR. Horizontal axis represents the varying number of observations.}\label{fig:families-coef}
\end{figure}

\end{knitrout}
Figure \ref{fig:families-eta} shows relative RIMSEs (left figure) and achieved point-wise coverages averaged across $t$ and $i$ (right figure, nominal 95\%) for the estimated additive predictor $\hat\eta_i(t)$ for each setting and replicate. Both point estimation and uncertainty quantification work well in these settings. For the point estimates, we observe the expected patterns of increasing precision with bigger data sets and decreasing noise levels. 
The systematic under-coverage for $n=300$, SNR$=5$ we observe for the Beta-settings for the model with a time-constant nonlinear interaction effect (third column, first row) seems of little practical relevance -- in this setting, the point estimates are so close to the true effects and estimation uncertainty becomes so small that the CIs almost vanish, and in any case observed coverages mostly remain above 90\% for nominal 95\%.

Figure \ref{fig:families-coef} shows relative RIMSEs (left) and coverages (right) for the estimated effects in the three settings that include more than an intercept. Again, we observe the expected patterns of increasing precision with bigger data sets and decreasing noise levels. Estimates are generally fairly precise and most coverages are close to or greater than the nominal 95\%-level. Systematic under-coverage occurs for the scalar interaction effect (middle column) for the Beta and $t(3)$-distributions, but note again that given the high precision of the point estimates the fact that CIs tend to be too narrow is of little practical relevance. This could also indicate a basis specification that is too small to fit the true effect without approximation bias. We also observe a single replicate of setting \texttt{smoo} with $t(3)$-distributed errors with much larger rRIMSE and coverage around 20\% (bottom row, leftmost panels, rightmost boxplots), indicating that the robustification by specifying a $t$-distribution may not always be successful.  

Median computation times for these fairly easy settings were between 3 and 49 seconds. Estimating the bivariate non-linear effect for setting \code{smoo} took the longest, with some fits taking up to 
8 minutes.
We observed no relevant or systematic differences in computation times between the three response distributions we used.

Appendix \ref{app:ex-families} shows some typical results for these settings and gives additional details on the data generating process as well as tabular representations of the performance measures in Figures \ref{fig:families-eta} and 
\ref{fig:families-coef}.

%% file: appendix_wang.tex
\subsection{Replication of Simulation Study in \citet{WangShi2014}}\label{app:wang}

This section describes our results of a replication of the simulation study described in Section 4.1 of \citet{WangShi2014}. We ran 20 replicates per setting.
The data comes from a logit model for binary data with just a functional intercept and observation-specific smooth functional random effects. The observation-specific smooth functional random effects $b_i(t)$ are realizations of a Gaussian process with ``squared exponential'' covariance structure,
the functional intercept $\beta_0(t)$ is a cubed sine-function, see \citet[their Section 4.1.]{WangShi2014} for details.
The model and data generating process are thus $P(y_i(t_l)=1) = \text{logit}^{-1}\left(\eta_i(t)\right)$ with $\eta_i(t) = \beta_0(t_l) + b_i(t_l)$; $i=1, \dots,n; l=1, \dots, T.$

Note that Wang \& Shi's MATLAB \citep{MATLAB} implementation of their proposal (denoted by \code{ggpfr}) uses the same ``squared exponential'' covariance structure for fitting as that used for generating the data, while our \code{pffr}-implementation uses a B-spline basis (8 basis functions per subject) to represent the $b_i(t)$. We used the same basis dimension for $\beta_0(t)$ for \code{ggpfr} as Wang \& Shi.

\begin{knitrout}\scriptsize
\definecolor{shadecolor}{rgb}{0.969, 0.969, 0.969}\color{fgcolor}\begin{figure}

{\centering \includegraphics[width=\textwidth]{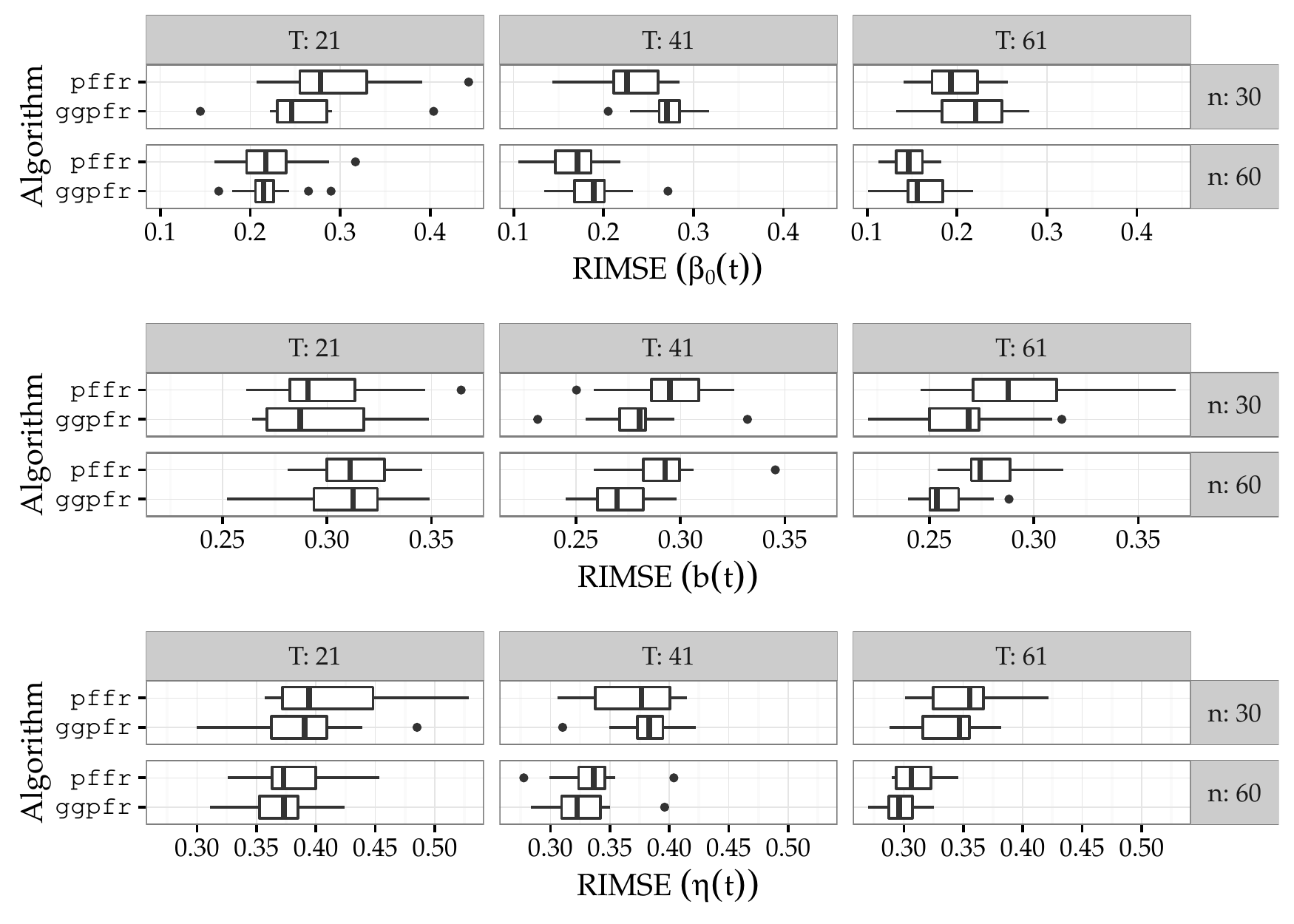} 

}

\caption{Replication of \citet{WangShi2014}: RIMSE for the estimated effects and the additive predictor. Top: functional intercept; middle: functional random effects; bottom: sum of the two. Each graph shows results for $n=30, 60$ (rows) and $T=21, 41, 61$ (columns). Upper boxplots for \code{pffr}, lower for \code{ggpfr}.}\label{fig:wang_rmse}
\end{figure}

\end{knitrout}

Boxplots in Figure \ref{fig:wang_rmse} show RIMSEs for the estimated effects and the additive predictor for each of the 20 replicates per combination of settings. While \code{ggpfr} tends to yield more precise estimates for $b_i(t)$, 
it often does not do as well as \code{pffr} in estimating the functional intercept.
As a consequence, the two approaches yield very similar errors for $\eta_i(t)$, even though the data generating process is tailored towards \code{ggpfr} and more adversarial for \code{pffr}. 
Note that we were not able to reproduce the mean RIMSE values for logit$(\eta(t))$ reported by \citet{WangShi2014} despite the authors' generous support. Our mean RIMSE results for \code{ggpfr} were $0.37, 0.33, 0.30$ while \citet{WangShi2014} reported $0.32, 0.26, 0.24$ for $n=60; T=21, 41, 61$,  respectively.

\begin{knitrout}\scriptsize
\definecolor{shadecolor}{rgb}{0.969, 0.969, 0.969}\color{fgcolor}\begin{figure}

{\centering \includegraphics[width=\textwidth]{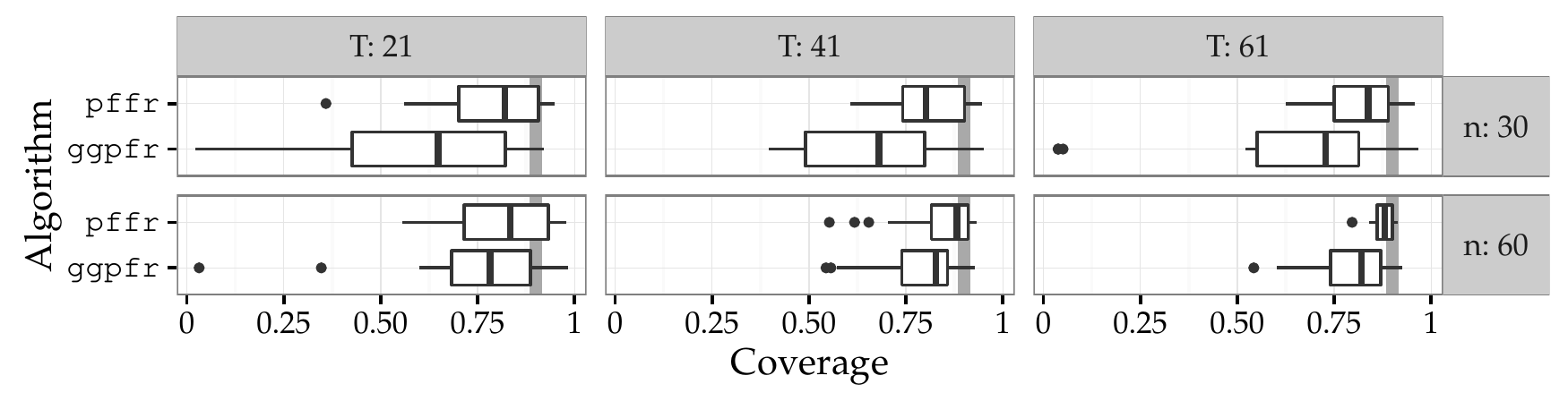} 

}

\caption{Replication of \citet{WangShi2014}: Observed CI coverage for $\eta(t) = \operatorname{logit}\left(\beta_0(t)+ b_i(t)\right)$ for $n=30, 60$ (rows) and $T=21, 41, 61$. Upper boxplots for \code{pffr}, lower for \code{ggpfr}. Vertical fat gray line denotes nominal 90\% coverage.}\label{fig:wang_cicover}
\end{figure}

\end{knitrout}
\begin{knitrout}\scriptsize
\definecolor{shadecolor}{rgb}{0.969, 0.969, 0.969}\color{fgcolor}\begin{figure}

{\centering \includegraphics[width=\textwidth]{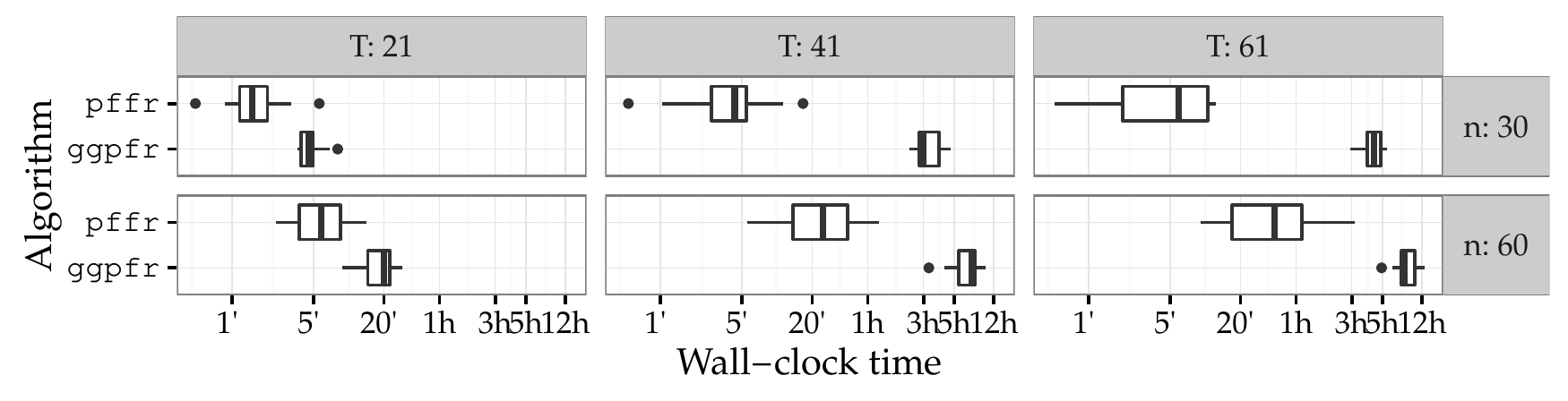} 

}

\caption{Replication of \citet{WangShi2014}: Computation times for $n=30, 60$ (rows) and $T=21, 41, 61$ (columns). Upper boxplots for \code{pffr}, lower for \code{ggpfr}. Horizontal axis on $\log_{10}$-scale.}\label{fig:wang_comptimes}
\end{figure}

\end{knitrout}
Figure \ref{fig:wang_cicover} shows observed CI coverage for nominal 90\% CIs for the two approaches evaluated over $\hat\eta_i(t)$. Both approaches mostly do not achieve nominal coverage for these very small data sets, but coverages for \code{pffr} converge towards the nominal level much faster as $n$ and $T$ increase than those for \code{ggpfr} and show fewer and smaller incidences of severe under-coverage.
Figure \ref{fig:wang_comptimes} shows computation times for the two approaches. 
The median computation time for \code{ggpfr} is about three times that of \code{pffr} for $T=21$, which increases to a factor of 13-19 for $n=60; T=41, 61$ and a factor of 40-48 for $n=30; T=41, 61$\footnote{Note that these computation times seem to be extremely sensitive towards different MATLAB versions -- in personal communication, Bo Wang reported average computation times of 3 minutes for $T=21, N=60$ (our result for \code{ggpfr}: $\approx 19$), 4 minutes for $T=41, N=30$ (our result: $\approx 200$) and 10 minutes for $T=61, N=60$ (our result: $\approx 540$) for a single run using MATLAB 7.4 on an Intel Core Duo CPU (2.53GHz, 3GB RAM) instead of MATLAB 7.12 which we used.}. Note that computation times for \code{ggpfr} are given for a single run using a pre-specified number of basis functions for estimating $\beta_0(t)$ here. But since the BIC-based selection of the basis dimension proposed by \citet{WangShi2014} requires $k$ runs for selecting between $k$ candidate numbers of basis functions, actual computation times for \code{ggpfr} in practical applications will increase roughly $k$-fold.

Our replication shows that \code{pffr} achieves mostly similar estimation accuracies with  better CI coverage for a data generating process tailored specifically to \code{ggpfr}'s strengths. Depending on the MATLAB version that is used, \code{pffr} does so in similar or much, much shorter time.

%% file: appendix_goldsmith.tex
\subsection{Replication of Simulation Study in \citet{Goldsmith2015}}\label{app:goldsmith}

This section describes our results of a replication of the simulation study described in Web Appendix 2 of \citet{Goldsmith2015}. We ran 20 replicates per setting. The data comes from a logit model for binary data with a functional intercept, a functional linear effect of a scalar covariate and observation-specific smooth functional random effects. The observation-specific smooth functional random effects $b_i(t)$ are drawn using 2 trigonometric functional principal component functions, the functional intercept $\beta_0(t)$ is a trigonometric function as well and the functional coefficient function $\beta_1(t)$ associated with a $N(0, 25)$-distributed scalar covariate $x$ is a scaled normal density function, see \citet[][Web Appendix, Section 2]{Goldsmith2015} for details. Note that \code{genfpca}, the \pkg{rstan} \citep{rstan} implementation provided by Goldsmith et al., uses the same number of FPCs (i.e, two) for the fit as used for generating the data, while our \code{pffr}-implementation uses 10 cubic B-spline basis functions to represent each $b_i(t)$.
The model and data generating process are thus $P(y_i(t_l)=1) = \text{logit}^{-1}\left(\eta_i(t)\right)$ with $\eta_i(t) = \beta_0(t_l) + \beta_1(t_l)x_i + b_i(t_l)$; $i=1, \dots,n; l=1, \dots, T.$

\begin{knitrout}\scriptsize
\definecolor{shadecolor}{rgb}{0.969, 0.969, 0.969}\color{fgcolor}\begin{figure}

{\centering \includegraphics[width=\textwidth]{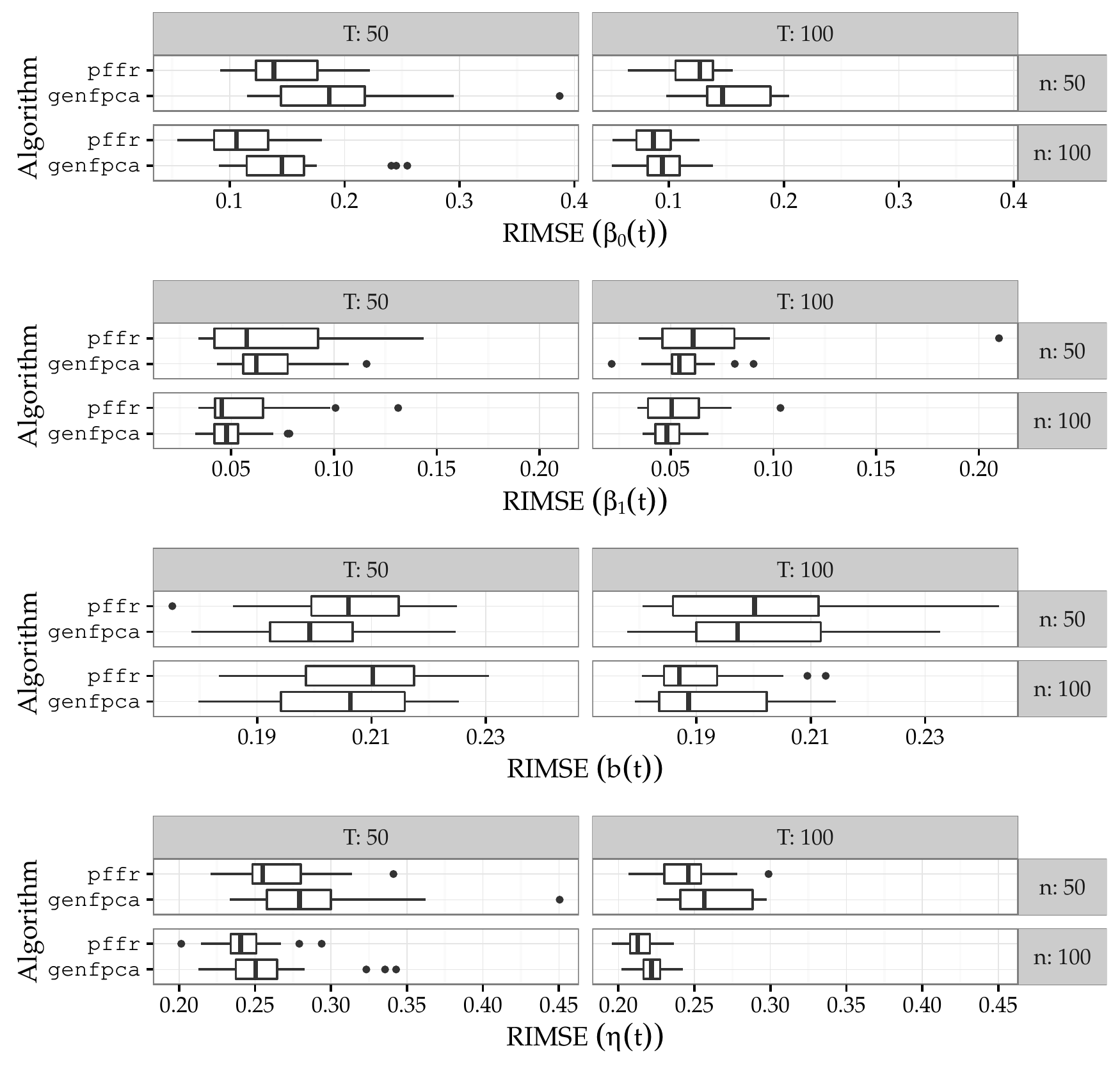} 

}

\caption{Replication of \citet{Goldsmith2015}: RIMSE for the estimated effects and the additive predictor. Top: functional intercept; 2nd row: functional coefficient; 3rd row: functional random effects; bottom: sum of the three. Each graph shows results for $n=50, 100$ (rows) and $T=50, 100$ (columns). Upper boxplots for \code{pffr}, lower for \code{genfpca}.}\label{fig:goldsmith_rmse}
\end{figure}

\end{knitrout}

\begin{knitrout}\scriptsize
\definecolor{shadecolor}{rgb}{0.969, 0.969, 0.969}\color{fgcolor}\begin{figure}

{\centering \includegraphics[width=\textwidth]{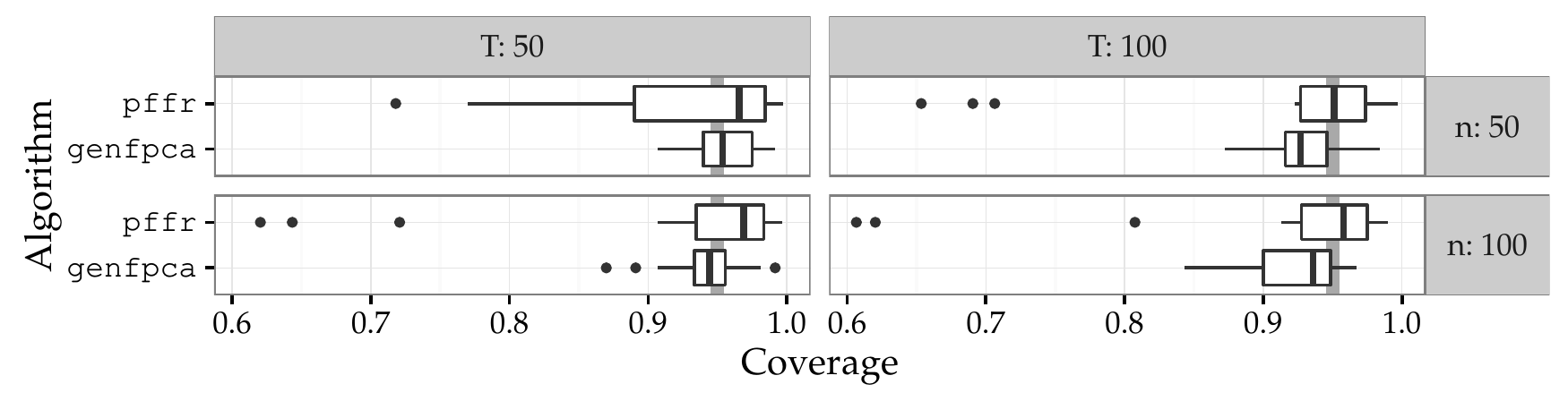} 

}

\caption{Replication of \citet{Goldsmith2015}: Observed CI coverage for $\eta_i(t) = \operatorname{logit}\left(\beta_0(t)+ \beta_1(t)x_i + b_i(t)\right)$ for $n=50, 100$ (rows) and $T=50, 100$. Upper boxplots for \code{pffr}, lower for \code{genfpca}. Vertical fat gray line denotes nominal 95\% coverage.}\label{fig:goldsmith_cicover}
\end{figure}

\end{knitrout}
\begin{knitrout}\scriptsize
\definecolor{shadecolor}{rgb}{0.969, 0.969, 0.969}\color{fgcolor}\begin{figure}

{\centering \includegraphics[width=\textwidth]{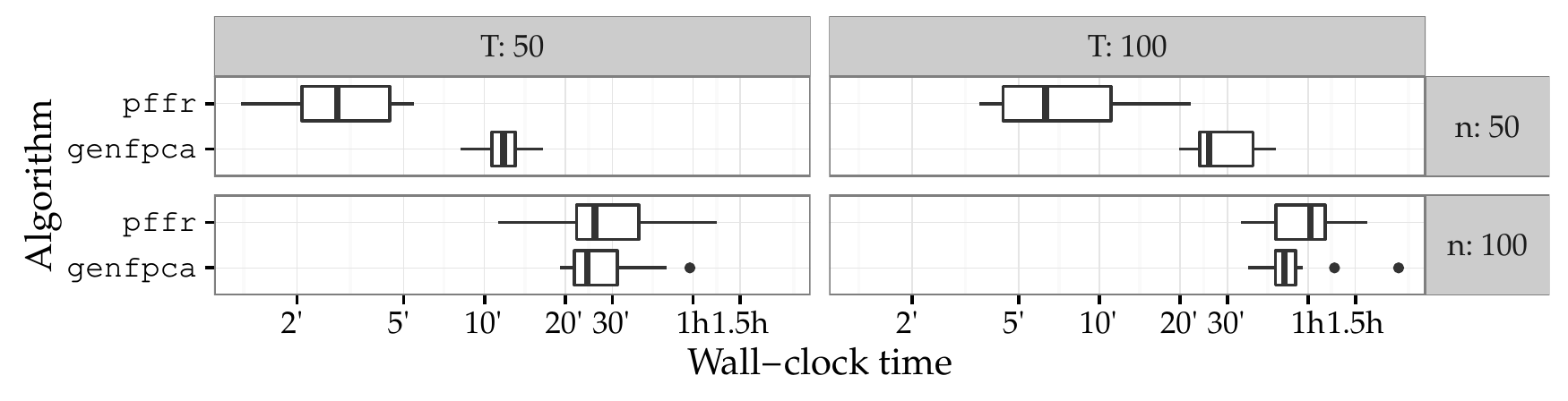} 

}

\caption{Replication of \citet{Goldsmith2015}: Computation times for $n=50, 100$ (rows) and $T=50, 100$ (columns). Upper boxplots for \code{pffr}, lower for \code{genfpca}. Horizontal axis on $\log_{10}$-scale.}\label{fig:goldsmith_comptimes}
\end{figure}

\end{knitrout}
Figure \ref{fig:goldsmith_rmse} shows RIMSE for (top to bottom) the estimated functional intercept, the estimated functional coefficient, the functional random effect and the combined additive predictor. It seems that \code{pffr} yields somewhat more precise estimates for $\beta_0(t)$ in this setting, while the estimation performance of the two approaches for the functional coefficient and the random effects is broadly similar although the results for \code{pffr} for $\beta_1(t)$ show more variability. 
\code{pffr} yields slightly better estimates of the total additive predictor $\eta_i(t)$ in this setting. Note that \citet{Goldsmith2015} report mean integrated square error (MISE, in their notation) while we report the RIMSE. Taking this into account, the \code{genfpca} results reported here correspond very closely to theirs. 
Figure \ref{fig:goldsmith_cicover} shows observed confidence [credibility] interval (CI) coverage for nominal 95\% CIs for the two approaches evaluated over $\hat\eta_i(t)$. Both approaches mostly achieve close to nominal coverage in this setting, but while coverages for \code{pffr} are mostly greater than the nominal $95$\% for $T=50$ with some rare problematic fits with coverages below 75\%, \code{genfpca} shows small but systematic under-coverage for $T=100$. Figure \ref{fig:goldsmith_comptimes} shows computation times for the two approaches. The median computation time for \code{genfpca} is about 4 times that of \code{pffr} for $n=50$, and $0.80$ to $0.94$ that of \code{pffr} for $n=100$.

In summary, our replication shows that \code{pffr} achieves mostly similar estimation accuracies and coverages in much shorter time or roughly the same time for a data generating process tailored specifically to \code{genfpca}'s strengths.

%% file: gfamm_discuss.tex
\section{Discussion}\label{sec:discuss}

This work introduces a comprehensive framework for generalized additive mixed models (GAMM) for non-Gaussian functional responses. Our implementation extends all the flexibility of GAMMs for dependent scalar responses to dependent functional responses and functional covariates, even for response distributions outside the exponential family such as robust models based on the $t$-distribution. Dependency structures can be spatial, temporal or hierarchical. Simulation and application results show that our approach provides reliable and precise inference about the underlying latent processes. 

The work presented here opens up promising new avenues of inquiry -- one challenge that we have already begun to work on is to improve the speed and memory efficiency of the underlying computational engine to be able to fit the huge data sets increasingly common in functional data analysis. Along these lines, more efficient computation of simultaneous instead of pointwise confidence intervals for functional effects as well as a more detailed investigation of the performance of the available approximate pointwise confidence intervals for noisy and small non-Gaussian data is another important field of inquiry.
An extension of the approach presented here to models with multiple additive predictors controlling different aspects of the responses' distribution -- like the generalized additive models for location, scale and shape introduced for scalar response by \citet{Rigby2005} or zero-inflation and hurdle models -- is yet another promising generalization of the ideas we have presented here.

%% file: gfamm_app.tex
\appendix

\pagenumbering{arabic}\renewcommand{\thepage}{\thesection.\arabic{page}}
\renewcommand\thefigure{\thesection.\arabic{figure}}
\setcounter{figure}{0}

\section{Additional Simulation Details}

Section \ref{app:ex-families} provides some more details on the data generating process used in Section \ref{sec:sim-families} and shows some typical effect estimates in these settings. 

\input{gfamm_ex_binomial.tex}


\input{appendix_ex_families.tex}

\subsection{Computational Details}

Code to reproduce results for Section 4 is included in a code supplement.\\
Results described here were computed on various Linux PCs and servers under \textsf{R-3.2.3} to \textsf{3.2.5} with \textsf{mgcv 1.8-5} to \textsf{1.8-7}, \textsf{refundDevel 0.1-11} to \textsf{0.1-15}.

\setcounter{figure}{0}

\input{gfamm_app_data.tex}

%% file: gfamm_ex_binomial.tex
\subsection{Simulation 1: Computation times, Details and Examples}\label{app:ex_binomial}

\subsubsection{Computation times}

\begin{knitrout}\scriptsize
\definecolor{shadecolor}{rgb}{0.969, 0.969, 0.969}\color{fgcolor}\begin{figure}

{\centering \includegraphics[width=\textwidth]{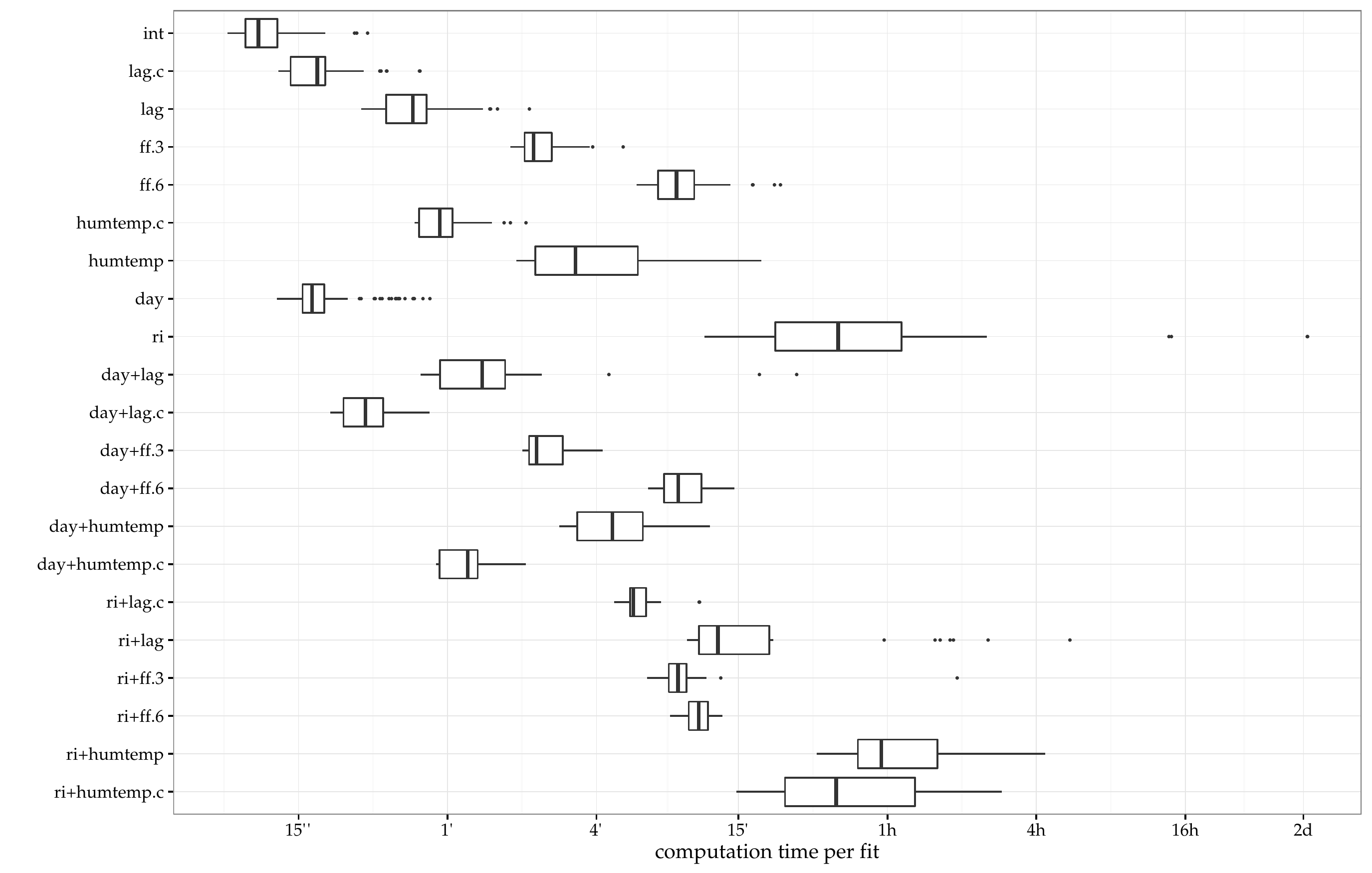} 

}

\caption[Computation times for the different models]{Computation times for the different models. Time axis on $\log_2$-scale.}\label{fig:binomial-time}
\end{figure}

\end{knitrout}
Figure \ref{fig:binomial-time} shows computation times for the various settings.
Note that the times are plotted on binary log-scale. One fit for a random intercept model (\code{ri}) did not converge within 200 iterations and ran for more than 16 hours.

\subsubsection{Computational details}

We use the following (spline) bases to construct $\mPhi_{\bm x}$  for the different effects:
\begin{itemize}
\item \texttt{lag.c, lag}: 20 cubic thin plate splines over $\tilde y(t)$
\item \texttt{humtemp, humtemp.c}: 50 bivariate cubic thin plate splines over the joint space of \texttt{hum} and \texttt{temp}.
\item \texttt{day}: 8 cubic thin plate splines over $i=1, \dots ,100$.
\item \texttt{ri}: dummy variables for each curve $i=1, \dots ,100$.
\end{itemize}
The coefficient surfaces for \texttt{ff.3, ff.6} are estimated with 30 bivariate cubic thin plate splines over $\mathcal T \times \mathcal T$. The functional intercept (\texttt{int}) uses 40 cubic cyclic P-splines with first order difference penalty over $\mathcal T$, while the functional random intercepts (\texttt{ri}) use 9 of those per curve.  For all other terms, we use 8 cubic P-splines with first order difference penalty for $\mPhi_{\bm t}$.

\subsubsection{Typical model fits}

Figures \ref{fig:binomial-daylag-ex} to \ref{fig:binomial-dayht-ex} show graphical summaries of typical model fits for three difficult settings of the simulated binomial data discussed in Section \ref{sec:sim-binomial}. 
It is clear to see that estimating unstructured daily functional random intercepts is a harder task than estimating a smoothly varying aging effect (compare the accuracy of $\hat f(t,i)$ in Figures \ref{fig:binomial-daylag-ex} and \ref{fig:binomial-dayht-ex} to that of $\hat b_{i}(t)$ in Figure \ref{fig:binomial-riff6-ex}), and that concurrent functional nonlinear interaction effects can not be recovered very reliably in this setting. Also note that we can model very rough latent response processes $\mu_i(t)$ for these data by including, e.g., a (nonlinear) auto-regressive term as in the top left panel of Figure \ref{fig:binomial-daylag-ex}.

\begin{knitrout}\scriptsize
\definecolor{shadecolor}{rgb}{0.969, 0.969, 0.969}\color{fgcolor}\begin{figure}

{\centering \includegraphics[width=\textwidth]{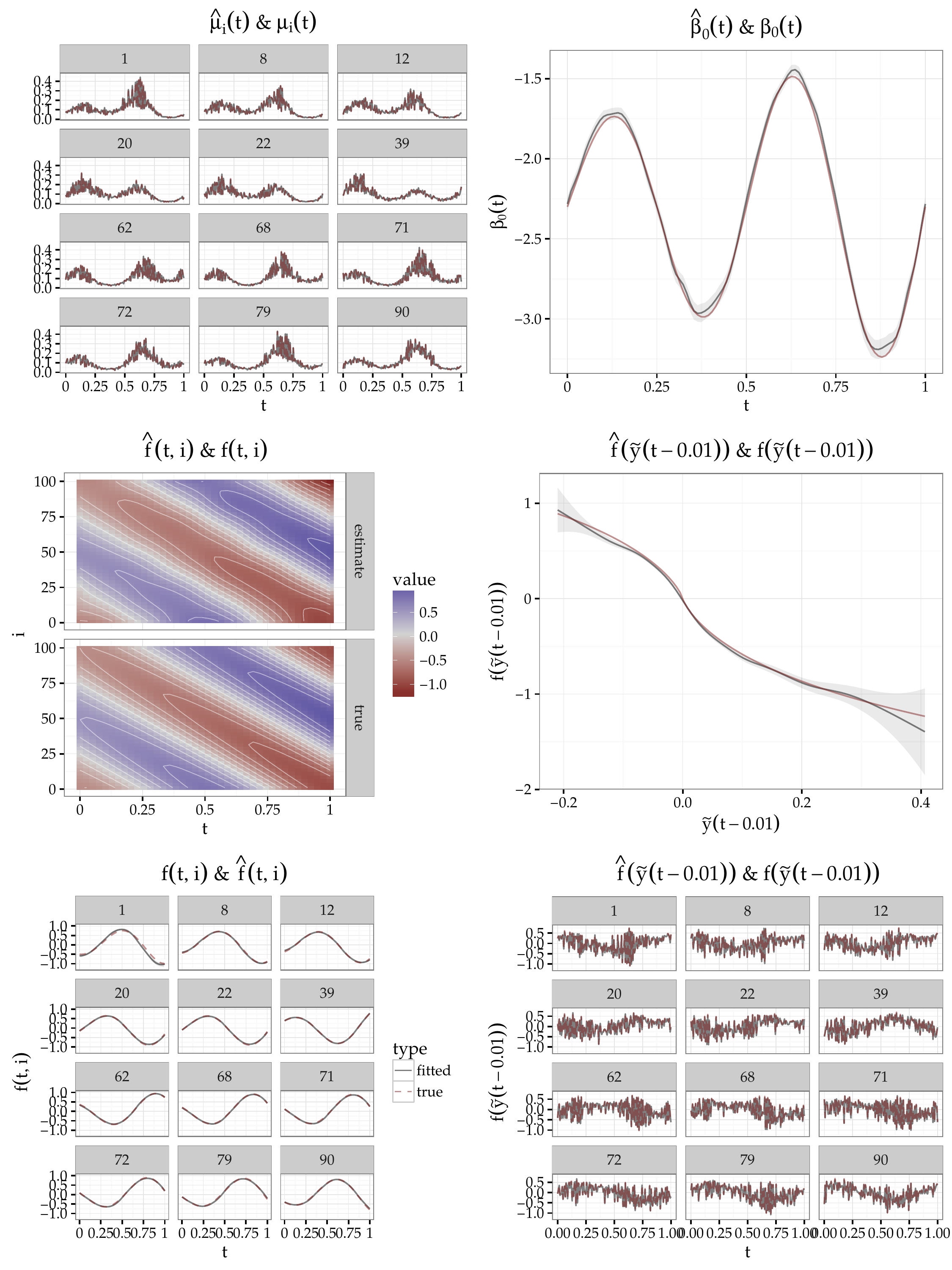} 

}

\caption{Typical model fit for setting \code{day+lag.c}: smoothly varying day effect with a nonlinear auto-regressive term.  Left column shows true (dashed, red) and estimated (solid, black) conditional expectation for 12 randomly selected days in the top row and estimated and true smoothly varying day-daytime effect in the middle and bottom panels. Right column shows true and estimated global intercept (top) and true and estimated nonlinear auto-regressive effect (middle) and its true and estimated contributions to the additive predictor. For this model, rRIMSE$(\hat\eta_i(t)) =  0.05 $, about the same as the median rRIMSE for this setting.}\label{fig:binomial-daylag-ex}
\end{figure}

\end{knitrout}

\begin{knitrout}\scriptsize
\definecolor{shadecolor}{rgb}{0.969, 0.969, 0.969}\color{fgcolor}\begin{figure}

{\centering \includegraphics[width=\textwidth]{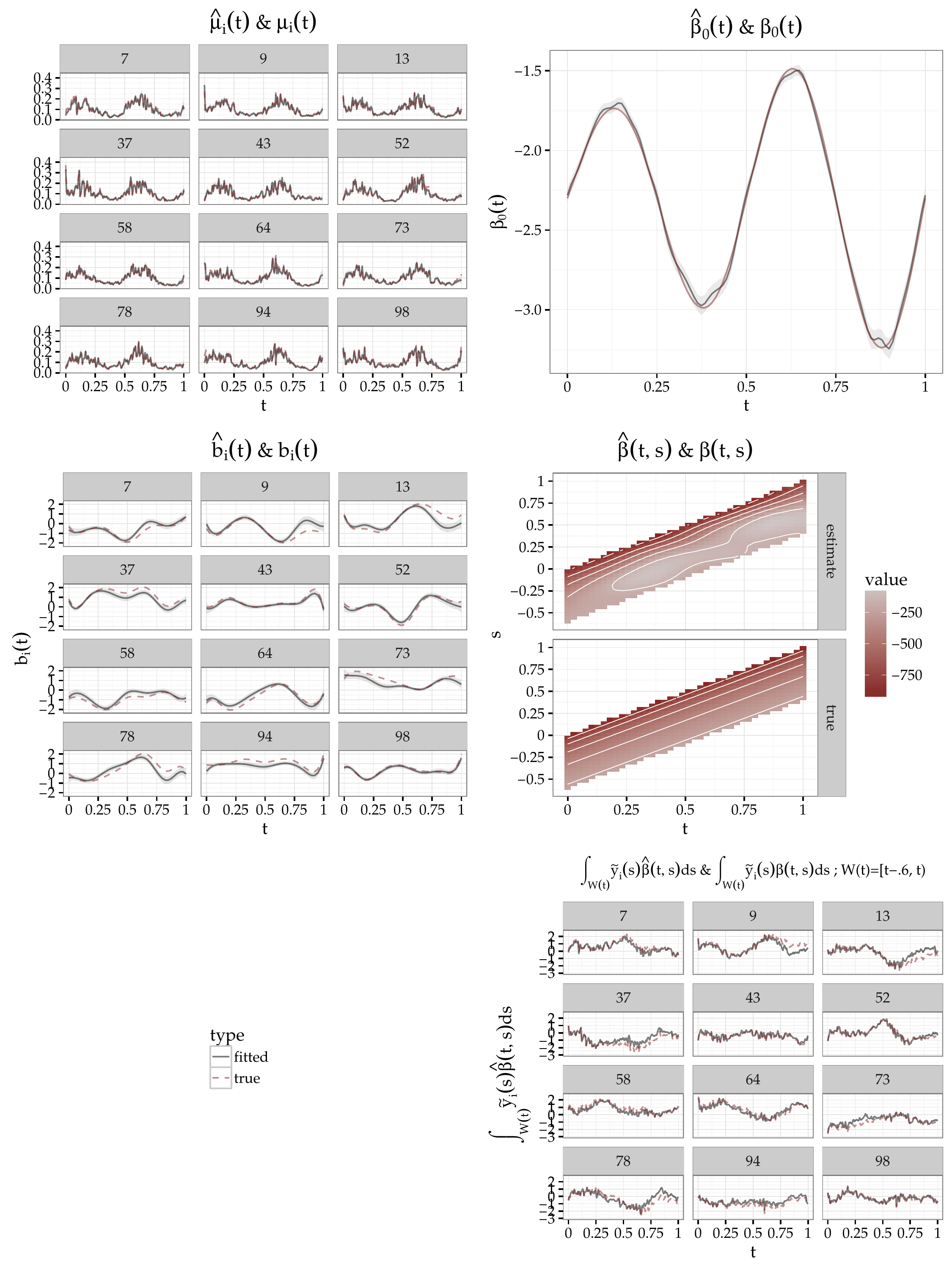} 

}

\caption{Typical model fit for setting \code{ri+ff.6}: random functional intercepts for each day with a cumulative auto-regressive term. Left column shows true (dashed, red) and estimated (solid, black) conditional expectation (top row) and estimated and true random intercepts (middle). Right column shows true and estimated global intercept (top) and estimated (middle, top panel) and true (middle, bottom panel) coefficient surface for the cumulative auto-regressive effect and true and estimated contributions of the cumulative auto-regressive effect to the additive predictor (bottom) for 12 randomly selected days. For this model, rRIMSE$(\hat\eta(t)) = 0.2$, while the median rRIMSE for this setting is 0.19.}\label{fig:binomial-riff6-ex}
\end{figure}

\end{knitrout}
\begin{knitrout}\scriptsize
\definecolor{shadecolor}{rgb}{0.969, 0.969, 0.969}\color{fgcolor}\begin{figure}

{\centering \includegraphics[width=\textwidth]{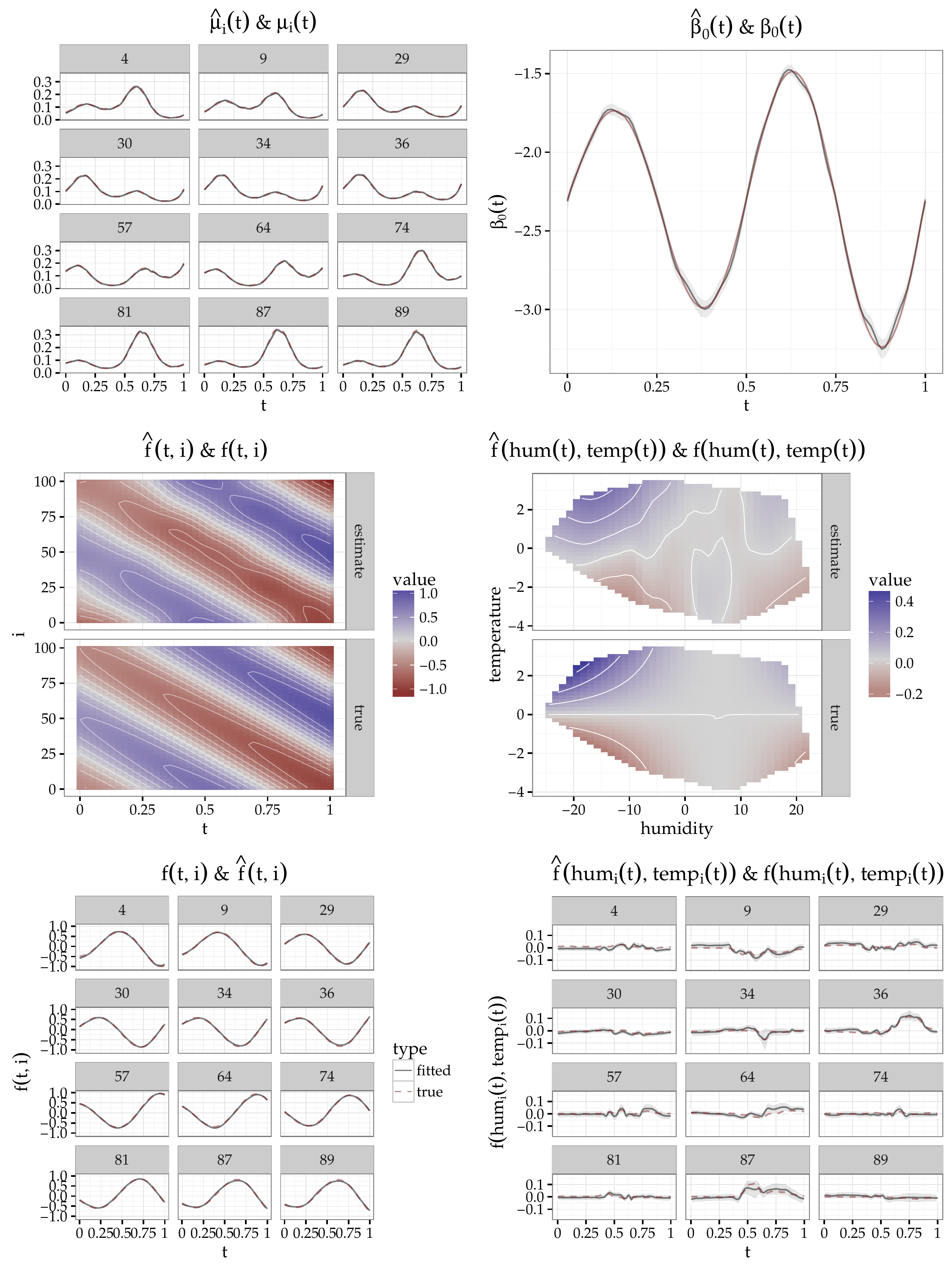} 

}

\caption{Typical model fit for setting \code{day+humtemp.c}: smoothly varying day effect with a time-constant concurrent nonlinear interaction effect of humidity and temperature. Left column shows true (dotted) and estimated (solid) conditional expectation (top) for 12 randomly selected days in the top row, estimated (top panel) and true (bottom panel) smoothly varying day-daytime effect (middle), and true and estimated contributions of the time-constant humidity-temperature interaction effect to the additive predictor (bottom). Right column shows true and estimated global intercept (top) and estimated (middle, top panel) and true (middle, bottom panel) effect surface for the humidity-temperature interaction. For this model, rRIMSE$(\hat\eta(t)) = 0.05$, while the median rRIMSE for this setting is 0.05.}\label{fig:binomial-dayht-ex}
\end{figure}

\end{knitrout}

%% file: appendix_ex_families.tex
\subsection{Simulation Study 2: Details \& Typical Fits}\label{app:ex-families}

\subsubsection{Data Generating Process for Simulation Study 2}

We use different signal-to-noise ratios (SNR) to control the difficulty of the simulation settings. For Beta-distributed responses with $g(\cdot)=\logit^{-1}(\cdot)$, distribution parameters $\alpha, \beta$ are determined by approximate SNR via $\phi = \text{SNR} \tfrac{m_\mu}{v_\mu}- 1$, with $m_\mu = N^{-1} \sum_{i,t} \mu_{it}(1-\mu_{it})$, where $v_\mu$ is the sample variance of the simulated $\mu_{it}$ and $\bm{\mu} = \text{logit}^{-1}(\meta)$. Beta parameters for generating $\my$ are then given by $\bm{\alpha} = \phi\bm{\mu}$ and  $\bm{\beta} = \phi - \bm{\alpha}$. $\meta$ is first scaled linearly to the interval $\left[-1.5, 1.5\right]$.

\subsubsection{Computational details}

We use the following spline bases to construct $\mPhi_{\bm x}$ for the different effects:
\begin{itemize}
\item \texttt{smoo}: 8 cubic thin plate splines over the range of $x$
\item \texttt{te}: 45 bivariate cubic thin plate splines over the joint space of $x_1$ and $x_2$
\item \texttt{ff}: 5 cubic P-splines with first order difference penalty over $\mathcal{S}$
\end{itemize}
The functional intercept (\texttt{int}) uses 40 cubic cyclic P-splines with first order difference penalty over $\mathcal T$. For all other terms, we use 5 cubic P-splines with first order difference penalty for $\mPhi_{\bm t}$.

\subsubsection{Typical Fits for Simulation Study 2}

Figures \ref{fig:families-ex-beta} to \ref{fig:families-ex-t3} show some typical fits for these data generating processes. Figure \ref{fig:families-ex-beta} shows estimated effects for a Beta-response model with a nonlinear effect of a scalar covariate (\code{smoo}, $n=300, \text{SNR}=1$), Figure \ref{fig:families-ex-nb} shows estimated effects for 
a Negative Binomial-response model with a linear function-on-function effect (\code{ff}, $n=100$) and Figure \ref{fig:families-ex-t3} shows estimated effects for 
a $t(3)$-response model with a nonlinear interaction effect of two scalar covariates (\code{te}, $n=100, \text{SNR}=5$)

\begin{knitrout}\scriptsize
\definecolor{shadecolor}{rgb}{0.969, 0.969, 0.969}\color{fgcolor}\begin{figure}

{\centering \includegraphics[width=\textwidth]{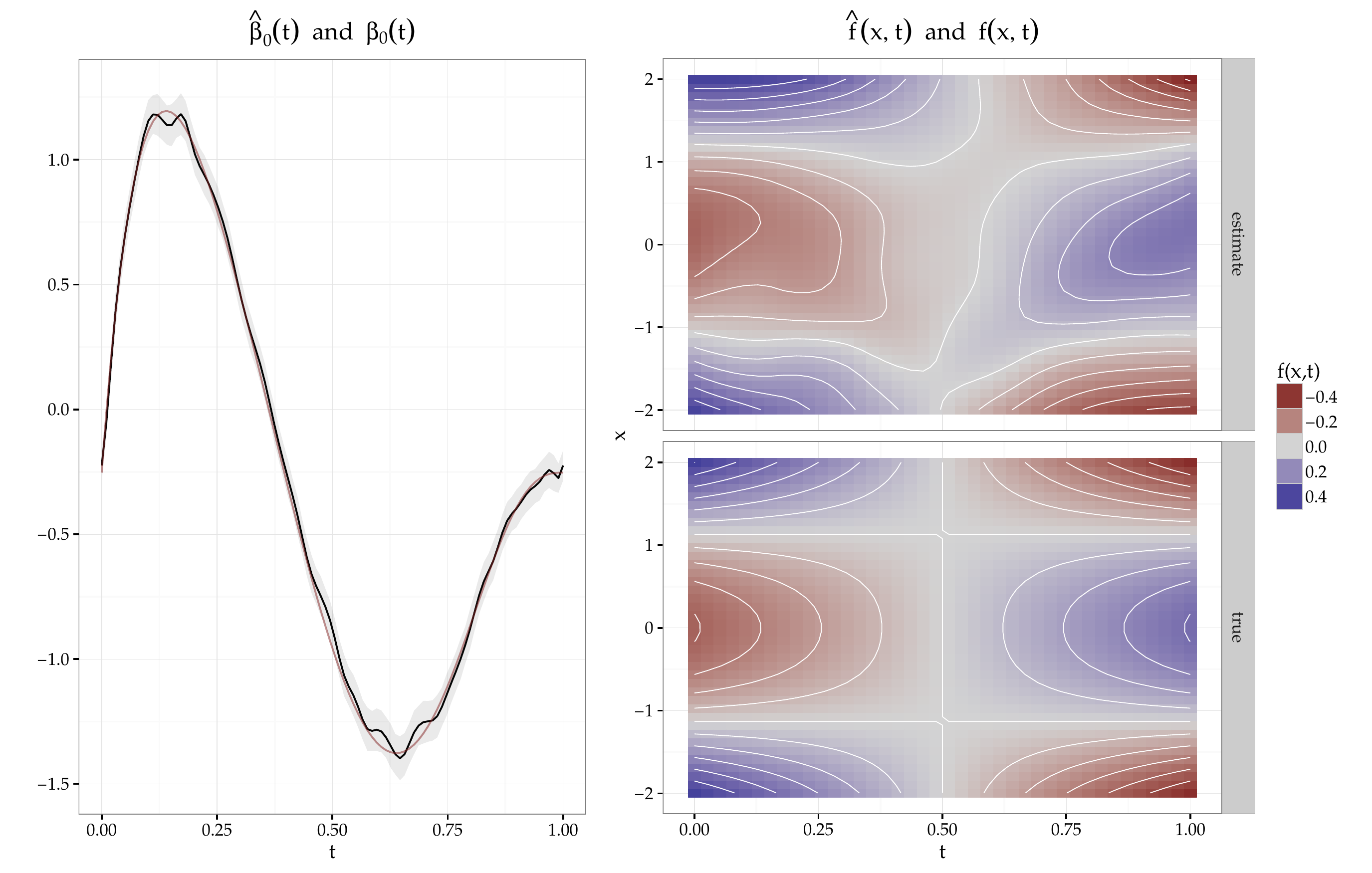} 

}

\caption[Typical  model  fit for a Beta-distributed model with a nonlinear effect of scalar covariate (\code{smoo}, ]{Typical  model  fit for a Beta-distributed model with a nonlinear effect of scalar covariate (\code{smoo}, $n=300, \text{SNR}=1$) with rRIMSE$(\hat\eta(t))\approx 0.047$ and coverage $\approx 0.98$ for $\hat\eta(t)$. Red line in the left panel shows true intercept function, light grey ribbon gives approximate pointwise 95\% interval. On the right, top panel shows estimated nonlinear effect of scalar covariate $\hat f(x,t)$ and bottom panel the true $f(x,t)$.}\label{fig:families-ex-beta}
\end{figure}

\end{knitrout}

\begin{knitrout}\scriptsize
\definecolor{shadecolor}{rgb}{0.969, 0.969, 0.969}\color{fgcolor}\begin{figure}

{\centering \includegraphics[width=\textwidth]{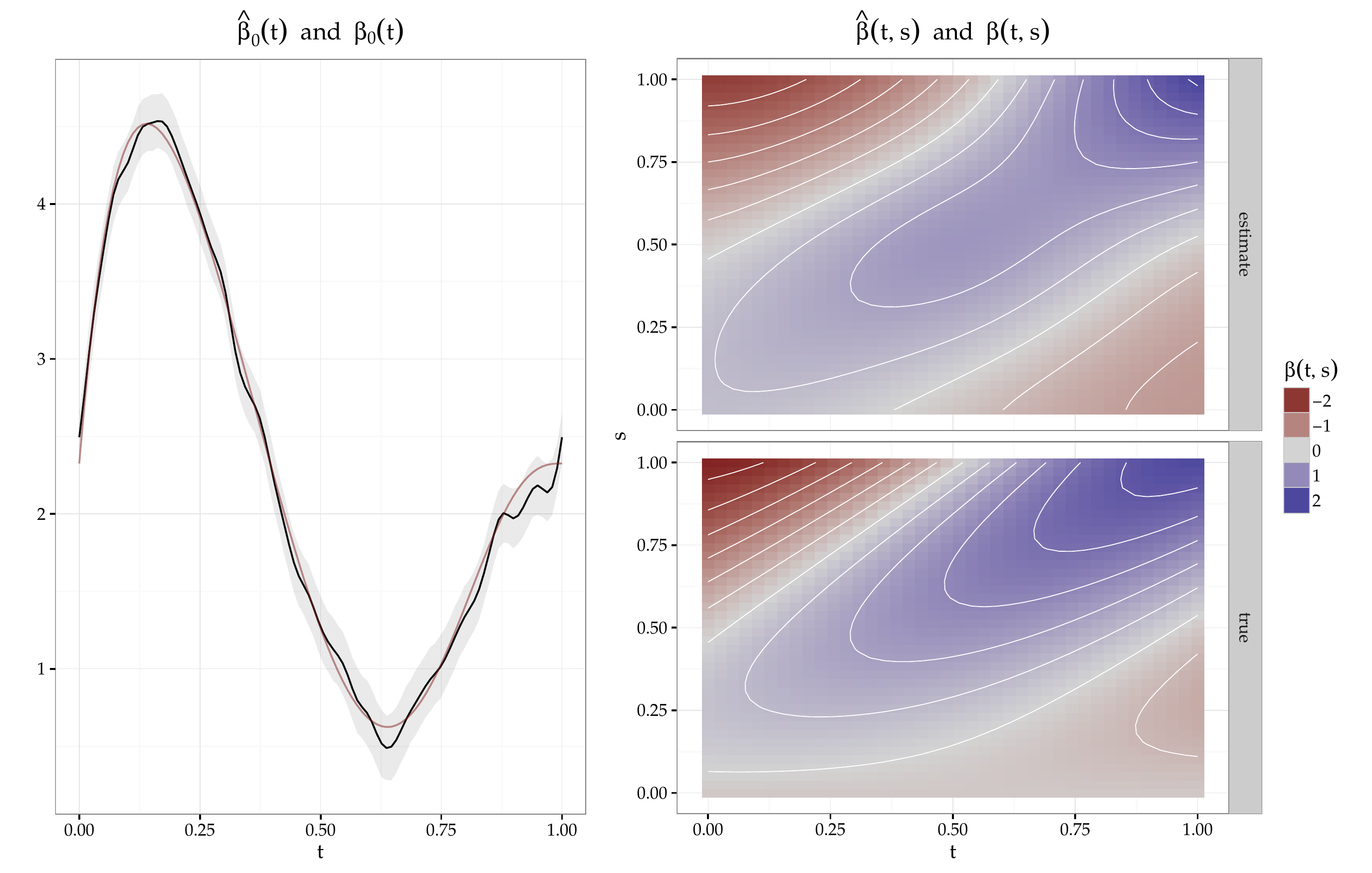} 

}

\caption[Typical  model  fit for a NB-distributed model with a linear effect of a functional covariate (\code{ff}, ]{Typical  model  fit for a NB-distributed model with a linear effect of a functional covariate (\code{ff}, $n=100$) with rRIMSE$(\hat\eta(t))\approx 0.083$ and coverage $\approx 0.97$ for $\hat\eta(t)$. Red dotted line in the left panel shows true intercept function, light grey ribbon gives approximate pointwise 95\% interval. On the right, top panel shows estimated $\hat \beta(s, t)$ and bottom panel the true  $\beta(s, t)$ for the function-on-function effect.}\label{fig:families-ex-nb}
\end{figure}

\end{knitrout}

\begin{knitrout}\scriptsize
\definecolor{shadecolor}{rgb}{0.969, 0.969, 0.969}\color{fgcolor}\begin{figure}

{\centering \includegraphics[width=\textwidth]{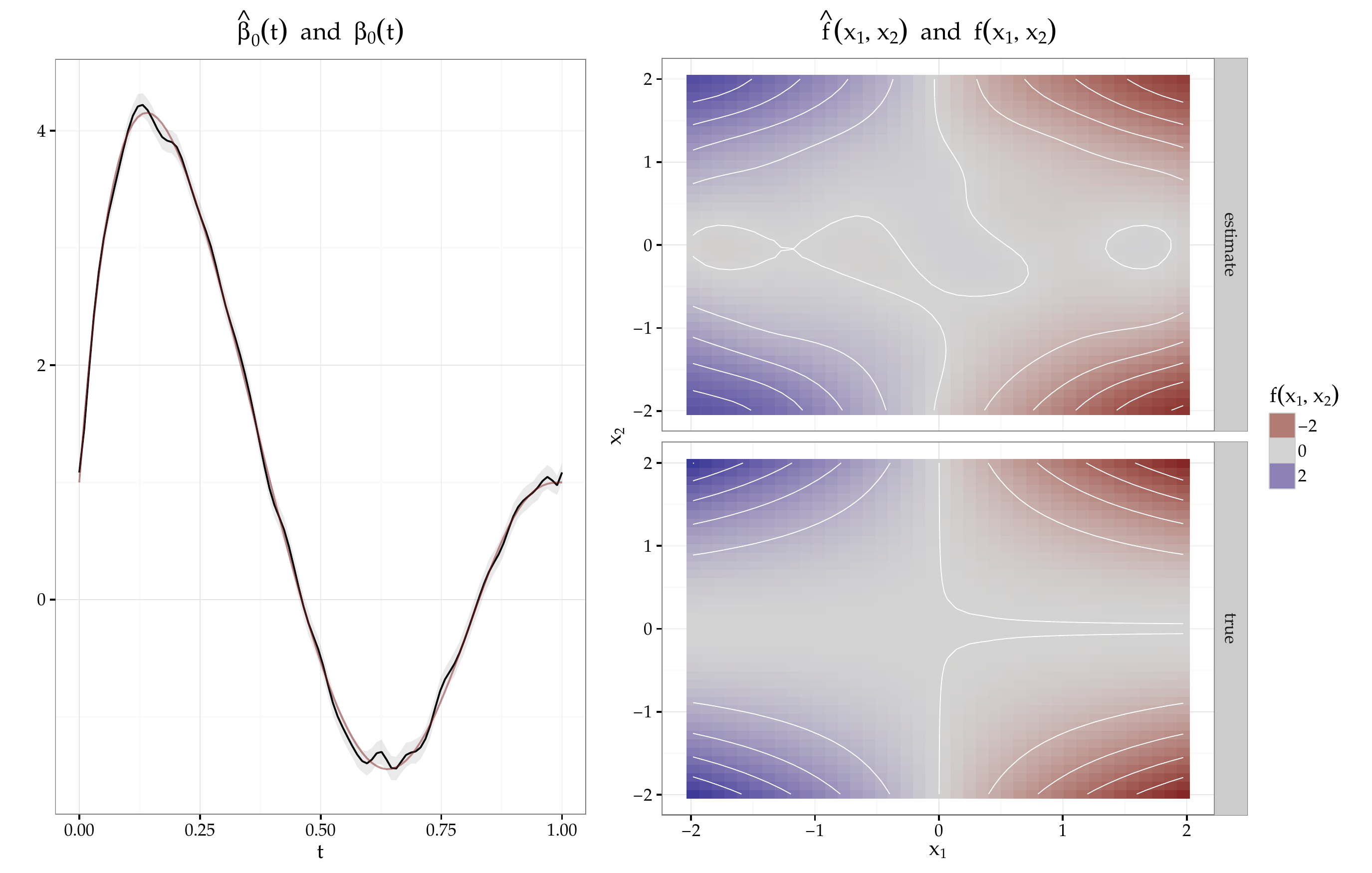} 

}

\caption[Typical model fit for a t3-distributed model with a nonlinear interaction effect of two scalar covariates (\code{te}, ]{Typical model fit for a t3-distributed model with a nonlinear interaction effect of two scalar covariates (\code{te}, $n=300, \text{SNR}=5$) with rRIMSE$(\hat\eta(t))\approx 0.043$ and coverage $\approx 0.94$ for $\hat\eta(t)$. Red dotted line in the left panel shows true intercept function, light grey ribbon gives approximate pointwise 95\% interval. On the right, top panel shows estimated smooth interaction $\hat f(x_1, x_2)$ and bottom panel the true  $f(x_1, x_2)$.}\label{fig:families-ex-t3}
\end{figure}

\end{knitrout}

\subsubsection{Tabular Results for Simulation Study 2}

Tables \ref{tab:families-tables-eta} and \ref{tab:families-tables-coef} show median
rRIMSES and coverages for simulation study \ref{sec:sim-families} for the entire additive predictor and the estimated covariate effects, respectively.

\begin{table}[ht]
\begin{footnotesize}
\begin{tabular}{llll|rrr|rrr}
  \hline
Family & Set & SNR & n & Median rRIMSE$(\hat\eta(t))$ & $q_{25}$ & $q_{75}$ & Median coverage & $q_{25}$ & $q_{75}$ \\ 
  \hline
Beta & int & 1 & 100 & 0.06 & 0.05 & 0.06 & 0.97 & 0.95 & 0.98 \\ 
   &  &  & 300 & 0.04 & 0.04 & 0.04 & 0.97 & 0.93 & 0.98 \\ 
   &  & 5 & 100 & 0.03 & 0.03 & 0.04 & 0.96 & 0.93 & 0.98 \\ 
   &  &  & 300 & 0.02 & 0.02 & 0.02 & 0.94 & 0.92 & 0.97 \\ 
   & smoo & 1 & 100 & 0.08 & 0.07 & 0.08 & 0.98 & 0.96 & 0.99 \\ 
   &  &  & 300 & 0.05 & 0.05 & 0.06 & 0.97 & 0.95 & 0.98 \\ 
   &  & 5 & 100 & 0.04 & 0.04 & 0.05 & 0.97 & 0.96 & 0.98 \\ 
   &  &  & 300 & 0.03 & 0.03 & 0.03 & 0.96 & 0.95 & 0.97 \\ 
   & te & 1 & 100 & 0.10 & 0.09 & 0.11 & 0.97 & 0.96 & 0.98 \\ 
   &  &  & 300 & 0.07 & 0.06 & 0.07 & 0.96 & 0.95 & 0.97 \\ 
   &  & 5 & 100 & 0.05 & 0.05 & 0.06 & 0.95 & 0.94 & 0.96 \\ 
   &  &  & 300 & 0.04 & 0.04 & 0.04 & 0.93 & 0.91 & 0.94 \\ 
   & ff & 1 & 100 & 0.07 & 0.07 & 0.08 & 0.96 & 0.94 & 0.98 \\ 
   &  &  & 300 & 0.05 & 0.05 & 0.05 & 0.96 & 0.94 & 0.97 \\ 
   &  & 5 & 100 & 0.04 & 0.04 & 0.04 & 0.96 & 0.94 & 0.97 \\ 
   &  &  & 300 & 0.03 & 0.02 & 0.03 & 0.95 & 0.93 & 0.96 \\ 
  NB & int & NA & 100 & 0.05 & 0.05 & 0.06 & 0.97 & 0.95 & 0.98 \\ 
   &  &  & 300 & 0.04 & 0.03 & 0.04 & 0.95 & 0.93 & 0.97 \\ 
   & smoo &  & 100 & 0.08 & 0.07 & 0.09 & 0.97 & 0.96 & 0.98 \\ 
   &  &  & 300 & 0.05 & 0.05 & 0.06 & 0.97 & 0.96 & 0.98 \\ 
   & te &  & 100 & 0.14 & 0.13 & 0.15 & 0.97 & 0.96 & 0.98 \\ 
   &  &  & 300 & 0.10 & 0.09 & 0.11 & 0.97 & 0.95 & 0.98 \\ 
   & ff &  & 100 & 0.08 & 0.08 & 0.09 & 0.96 & 0.94 & 0.97 \\ 
   &  &  & 300 & 0.05 & 0.05 & 0.06 & 0.95 & 0.94 & 0.97 \\ 
  t(3) & int & 1 & 100 & 0.07 & 0.06 & 0.08 & 0.97 & 0.95 & 0.98 \\ 
   &  &  & 300 & 0.05 & 0.04 & 0.05 & 0.97 & 0.95 & 0.98 \\ 
   &  & 5 & 100 & 0.04 & 0.03 & 0.04 & 0.95 & 0.93 & 0.97 \\ 
   &  &  & 300 & 0.02 & 0.02 & 0.03 & 0.95 & 0.92 & 0.97 \\ 
   & smoo & 1 & 100 & 0.09 & 0.08 & 0.10 & 0.98 & 0.96 & 0.99 \\ 
   &  &  & 300 & 0.06 & 0.05 & 0.06 & 0.97 & 0.96 & 0.98 \\ 
   &  & 5 & 100 & 0.05 & 0.05 & 0.05 & 0.97 & 0.96 & 0.98 \\ 
   &  &  & 300 & 0.03 & 0.03 & 0.03 & 0.97 & 0.95 & 0.97 \\ 
   & te & 1 & 100 & 0.12 & 0.11 & 0.12 & 0.97 & 0.96 & 0.98 \\ 
   &  &  & 300 & 0.08 & 0.07 & 0.08 & 0.96 & 0.95 & 0.97 \\ 
   &  & 5 & 100 & 0.06 & 0.06 & 0.07 & 0.95 & 0.94 & 0.97 \\ 
   &  &  & 300 & 0.04 & 0.04 & 0.04 & 0.94 & 0.93 & 0.95 \\ 
   & ff & 1 & 100 & 0.09 & 0.08 & 0.09 & 0.96 & 0.94 & 0.98 \\ 
   &  &  & 300 & 0.06 & 0.05 & 0.06 & 0.96 & 0.94 & 0.97 \\ 
   &  & 5 & 100 & 0.05 & 0.04 & 0.05 & 0.96 & 0.94 & 0.97 \\ 
   &  &  & 300 & 0.03 & 0.03 & 0.03 & 0.95 & 0.93 & 0.96 \\ 
   \hline
\end{tabular}
\caption{Tabular display of results in Figure \ref{fig:families-eta}. Median and 25\% and 75\% quantiles for relative RIMSE and pointwise coverages for the additive predictor $\hat\eta(t)$.} 
\label{tab:families-tables-eta}
\end{footnotesize}
\end{table}

\begin{table}[ht]
\begin{footnotesize}
\begin{tabular}{llll|rrr|rrr}
  \hline
Family & Set & SNR & n & Median rRIMSE$(\hat f(\mathcal{X}_{r}, t))$ & $q_{25}$ & $q_{75}$ & Median coverage & $q_{25}$ & $q_{75}$ \\ 
  \hline
Beta & smoo & 1 & 100 & 0.29 & 0.25 & 0.33 & 0.99 & 0.97 & 1.00 \\ 
   &  &  & 300 & 0.20 & 0.18 & 0.22 & 0.99 & 0.97 & 1.00 \\ 
   &  & 5 & 100 & 0.17 & 0.16 & 0.19 & 0.99 & 0.97 & 1.00 \\ 
   &  &  & 300 & 0.12 & 0.11 & 0.13 & 0.98 & 0.96 & 0.99 \\ 
   & te & 1 & 100 & 0.18 & 0.16 & 0.20 & 0.94 & 0.92 & 0.96 \\ 
   &  &  & 300 & 0.11 & 0.10 & 0.12 & 0.94 & 0.92 & 0.96 \\ 
   &  & 5 & 100 & 0.13 & 0.11 & 0.15 & 0.90 & 0.88 & 0.92 \\ 
   &  &  & 300 & 0.07 & 0.07 & 0.08 & 0.89 & 0.87 & 0.91 \\ 
   & ff & 1 & 100 & 0.32 & 0.28 & 0.36 & 0.93 & 0.87 & 0.97 \\ 
   &  &  & 300 & 0.20 & 0.18 & 0.23 & 0.94 & 0.90 & 0.99 \\ 
   &  & 5 & 100 & 0.17 & 0.15 & 0.19 & 0.95 & 0.90 & 0.99 \\ 
   &  &  & 300 & 0.11 & 0.10 & 0.12 & 0.94 & 0.91 & 0.98 \\ 
  NB & smoo & NA & 100 & 0.31 & 0.27 & 0.35 & 0.99 & 0.97 & 1.00 \\ 
   &  &  & 300 & 0.21 & 0.18 & 0.22 & 0.99 & 0.97 & 1.00 \\ 
   & te &  & 100 & 0.23 & 0.22 & 0.26 & 0.96 & 0.93 & 0.97 \\ 
   &  &  & 300 & 0.16 & 0.14 & 0.17 & 0.96 & 0.94 & 0.97 \\ 
   & ff &  & 100 & 0.35 & 0.31 & 0.40 & 0.93 & 0.85 & 0.97 \\ 
   &  &  & 300 & 0.22 & 0.19 & 0.24 & 0.94 & 0.89 & 0.97 \\ 
  t(3) & smoo & 1 & 100 & 0.32 & 0.29 & 0.36 & 1.00 & 0.98 & 1.00 \\ 
   &  &  & 300 & 0.22 & 0.20 & 0.25 & 0.99 & 0.97 & 1.00 \\ 
   &  & 5 & 100 & 0.19 & 0.18 & 0.21 & 0.99 & 0.97 & 0.99 \\ 
   &  &  & 300 & 0.13 & 0.12 & 0.14 & 0.98 & 0.96 & 0.99 \\ 
   & te & 1 & 100 & 0.20 & 0.18 & 0.22 & 0.95 & 0.93 & 0.97 \\ 
   &  &  & 300 & 0.12 & 0.11 & 0.13 & 0.95 & 0.94 & 0.97 \\ 
   &  & 5 & 100 & 0.14 & 0.12 & 0.15 & 0.91 & 0.89 & 0.93 \\ 
   &  &  & 300 & 0.08 & 0.07 & 0.09 & 0.90 & 0.88 & 0.92 \\ 
   & ff & 1 & 100 & 0.36 & 0.32 & 0.41 & 0.93 & 0.87 & 0.98 \\ 
   &  &  & 300 & 0.24 & 0.20 & 0.26 & 0.94 & 0.89 & 0.98 \\ 
   &  & 5 & 100 & 0.19 & 0.17 & 0.22 & 0.94 & 0.90 & 0.99 \\ 
   &  &  & 300 & 0.13 & 0.11 & 0.14 & 0.94 & 0.90 & 0.97 \\ 
   \hline
\end{tabular}
\caption{Tabular display of results in Figure \ref{fig:families-coef}.  Median and 25\% and 75\% quantiles for relative RIMSE and pointwise coverages for the estimated effects $\hat f(\mathcal{X}_{r}, t)$.} 
\label{tab:families-tables-coef}
\end{footnotesize}
\end{table}

%% file: gfamm_app_data.tex
\section{PIGWISE Model: Alternatives and Criticism\label{app:alternative-models}}

\begin{figure}
\begin{center}
\includegraphics[width=.8\textwidth]{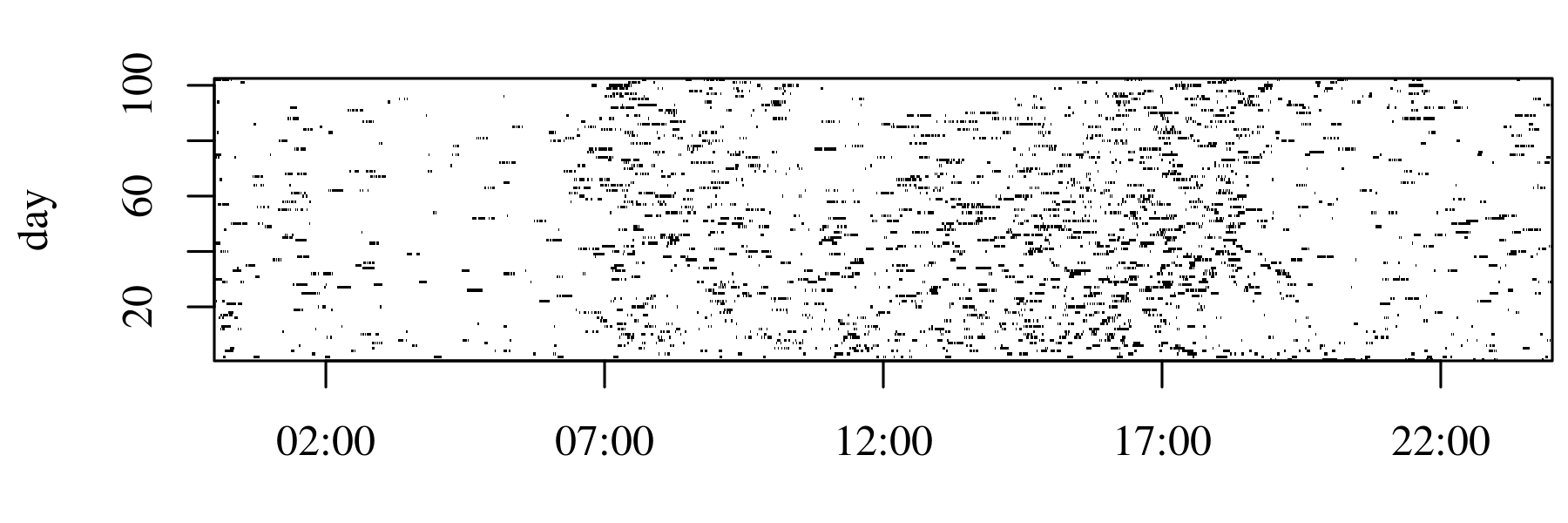}
\end{center}
\caption[Observed feeding episodes for pig 57]{Observed feeding episodes for pig 57. Horizontal axis gives time of day, vertical axis represents days. Black dots show observed feeding episodes as measured by proximity to the trough (yes-no).}\label{fig:piggyplot1}
\end{figure}

This section shows the raw data used for the PIGWISE application (Figure~\ref{fig:piggyplot1}) and compares the result of a different  model specification to that of the main article.

Beyond the smoothly varying functional day effects used in the analysis in Section 
\ref{sec:app-model}, we considered two alternative random effect spe\-ci\-fi\-cations:
auto-correlated functional random day effects with a marginal $AR(1)$-structure with
auto-correlation 0.8 over the days as well as $\iid$ random day effects $b_i(t)$. We also considered models with no day effects at all.

Beyond the 3h-cumulative auto-regressive effect of previous feeding used in the analysis in Section \ref{sec:app-model}, we also fit  models with a 6h-cumulative auto-regressive effect
($\int^{t-10\text{min}}_{t-6\text{h}} y_i(s) \beta(t,s)ds$) as well as non-linear time-constant auto-regressive effects ($f(y_i(t-10\text{min}))$) and time-varying ($f(y_i(t-10\text{min}), t)$). We also considered models with no auto-regressive effects at all. 

To model possible effects of humidity and temperature, we considered additive non-linear time-varying ($f(\text{hum}(t), t) + f(\text{temp}(t), t)$) and 
time-constant ($f(\text{hum}(t)) + f(\text{temp}(t))$) concurrent effects, as well as
corresponding concurrent interaction effects  $f(\text{hum}(t), \text{temp}(t))$ and $f(\text{hum}(t), \text{temp}(t), t)$, respectively. We also considered models with no 
effects of humidity and/or temperature at all. 

We estimated models for all 100 combinations (4 day effects, 5 auto-regressive effects, 5 humidity/temperature effects) of the different effect specifications given above.  As in the main analysis, we excluded every third day from the training sample to serve as a validation set. Analysis of mean Brier scores achieved on the training data showed that all 100 models fit the training data similarly well, with slightly better fits for models with $\iid$ or $AR(1)$ functional random effects compared to models with no day effects or a smooth day effect.
Analysis of mean Brier scores on the validation data, however, showed that only very small differences in predictive accuracy between models with a smooth day effect and no day effects at all exist and also revealed a strong decrease in predictive performance for models with humidity-temperature effects, especially in combination with cumulative auto-regressive effects.
All in all, the majority of models performed worse in terms of predictive Brier score than a simple functional intercept model $y_i(t)  \sim B(60, \text{logit}^{-1}(\beta_0(t)))$, which had a mean predictive Brier score of 0.026, compared to a mean predictive Brier score of 0.029 for the $\iid$ functional random intercept model discussed below and a mean predictive Brier score of 0.024 of a model like the one discussed in Section \ref{sec:app-res} with a smoothly varying day effect better suited for generating interpolating predictions for missing days instead of functional random day effects, which are constant 0 for days in the test set. We conclude that overfitting of complex effects could be a serious concern for the model class we propose, as shown by the large differences in Brier scores between training and validation data for many of the models with complicated effect structures. Note, however, that even though the $\iid$ random effect model yields inferior predictions, it may be better suited for estimating explanatory and interpretable models than the one discussed in Section \ref{sec:app-model}, as the larger flexibility of the $\iid$ functional random  day effects is presumably more effective at modeling the peaky and irregular temporal dependencies in the observations than a smoothly varying aging effect. 

\begin{knitrout}\scriptsize
\definecolor{shadecolor}{rgb}{0.969, 0.969, 0.969}\color{fgcolor}\begin{figure}

{\centering \includegraphics[width=\textwidth]{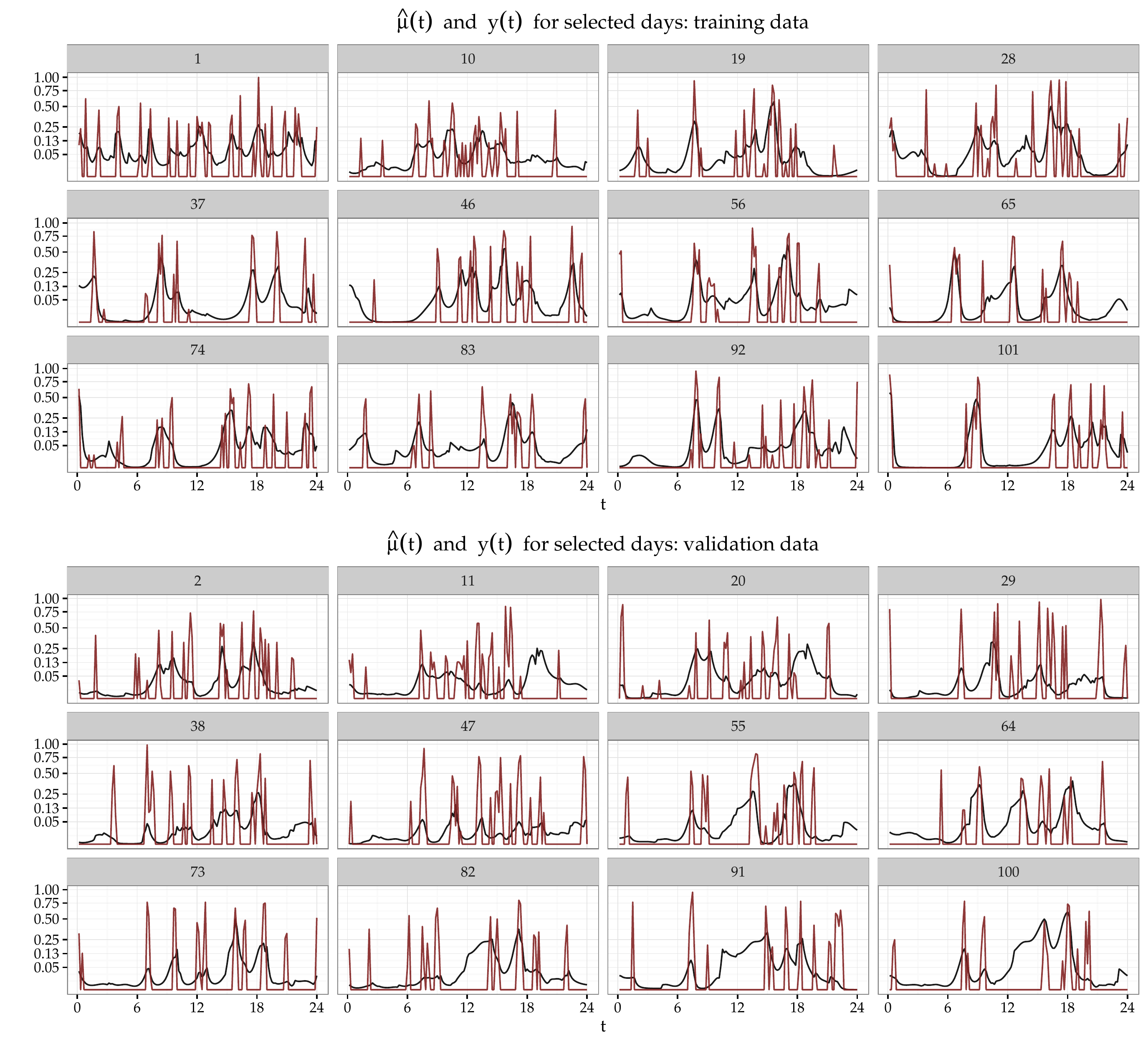} 

}

\caption{Fitted (top) and predicted (bottom) values for pig 57 for selected days for an alternative model specification with $\iid$ functional random day effects instead of a smooth aging effect. Black lines for fitted (top) and predicted (bottom) values $\hat\mu_i(t)$, red for observed feeding rate $y_i(t)/60$. Numbers above each panel give the day. Vertical axis on $\sqrt{\;}$-scale.}\label{fig:plot_m62_fit}
\end{figure}

\end{knitrout}

We present detailed results for an alternative model specification with $\iid$ functional random day effects $b_i(t)$ instead of a smooth aging effect $f(i, t)$ below.  Figure \ref{fig:plot_m62_fit} shows fitted and observed values for 12 selected days from the training data (top) as well as predicted and observed values for 12 selected days from the validation data (bottom). It is easy to see that the model is able to reproduce many of the feeding episodes in the training data, in the sense that peaks in the estimated probability curves mostly line up well with observed spikes of $y_i(t)$. This model explains about 41\% of the deviance and achieves a Brier score of about 0.019 taking the mean over all days and time-points. Prediction (lower panel) with this model is more challenging (Brier score: 0.029), and succeeds only partially. For example, while the predictions for day 73 are mostly very good, predictions for the evening hours of day 11 or around noon on day 82 are far off the mark. Note, however, that $\hat b_i(t) \equiv 0$ for the validation data, as $E(b_i(t)) \equiv 0 \,\forall\, t, i.$ Models that include a smooth day effect $f(i, t)$ instead, which can be interpolated for the days in the calibration data, are more successful at prediction, see Section \ref{sec:app-res}. 

\begin{knitrout}\scriptsize
\definecolor{shadecolor}{rgb}{0.969, 0.969, 0.969}\color{fgcolor}\begin{figure}

{\centering \includegraphics[width=\textwidth]{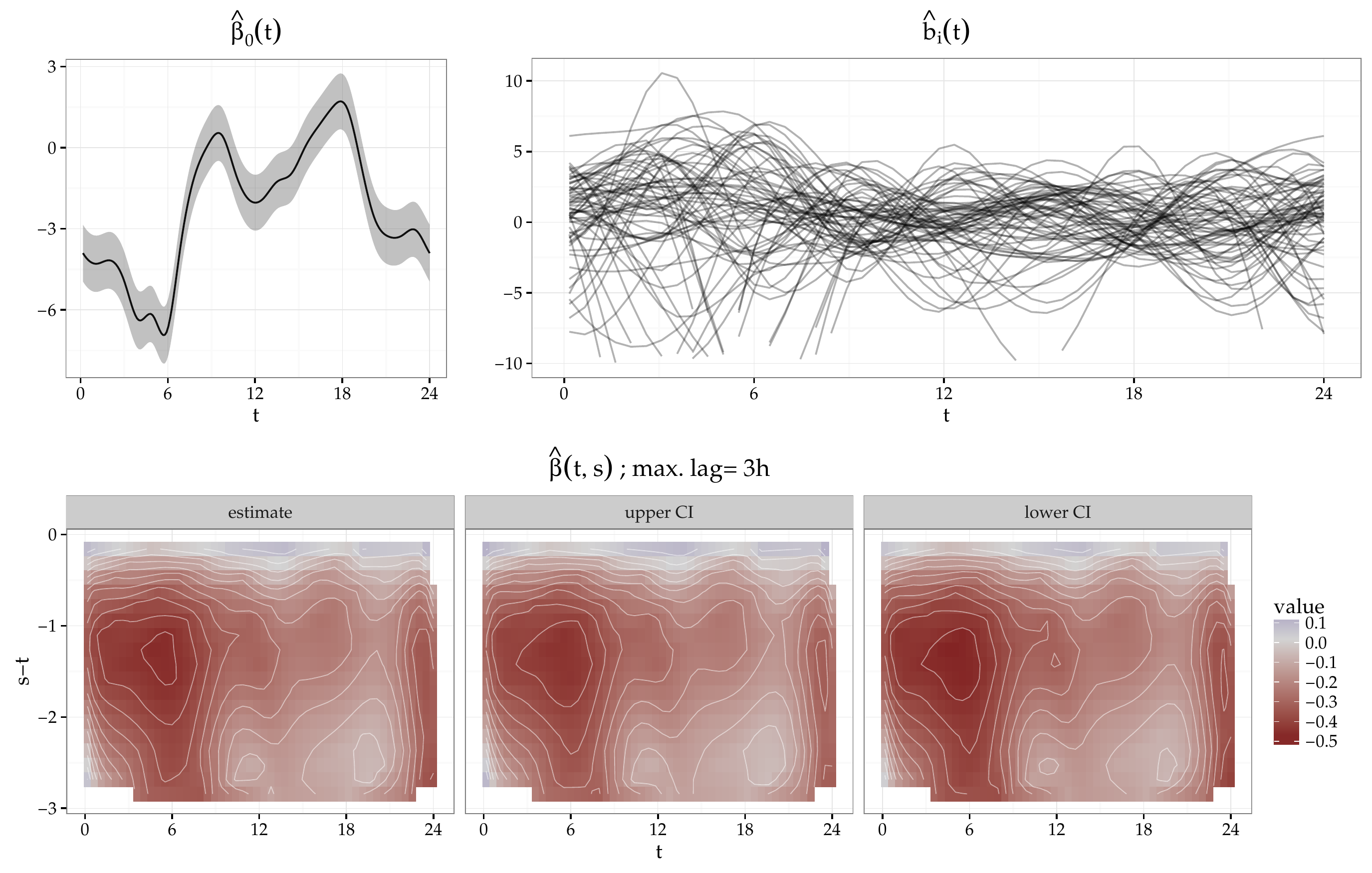} 

}

\caption[Estimated effects for an alternative model specification with ]{Estimated effects for an alternative model specification with $\iid$ functional random day effects instead of a smooth aging effect. Top row: Estimated functional intercept and functional random day effects. Bottom row: Estimated coefficient surface for cumulative auto-regressive effect. Intervals are  $\pm$ 2 standard errors. Vertical axis for functional random day effects is truncated at $-10$ to show structure of typical results.}\label{fig:plot_m62_coef}
\end{figure}

\end{knitrout}

Figure \ref{fig:plot_m62_coef} shows the estimated components of the additive predictor for this alternative model specification. 
The functional intercept in the top left panel of Figure \ref{fig:plot_m62_coef}
is fairly similar in shape to Figure \ref{fig:plot_m64_coef}, but overall the base rate is estimated to be much higher and the peak before 6h is much less pronounced.
Estimated random day effects are shown in the top right panel of Figure \ref{fig:plot_m62_coef}. In terms of absolute size, these are much larger than the values of the smooth aging effect depicted in Figure \ref{fig:plot_m64_coef}. Effect sizes for $b_i(t)$ are often unrealistically large, with some estimates going as low as -20 (-20 on the logit scale corresponds to a reduction of the probability by a factor of about $2 \cdot 10^{-9}$), presumably an artifact of the very low feeding activity between midnight and early morning leading to (quasi-)complete separability of the data. The vertical axis in Figure \ref{fig:plot_m62_coef} is cut off at $-10$ in order to showcase the typical ``peaky'' structure of the $\hat b_i(t)$ which would otherwise be hidden by the larger vertical axis required to show the $12$ random intercept functions reaching minimal values below $-10$ in their entirety.
The cumulative auto-regressive effect of feeding during the previous 3 hours displayed in the bottom row of Figure \ref{fig:plot_m62_coef} is quite similar to the estimate for the original model shown in Figure \ref{fig:plot_m64_coef} but with a weaker positive association between feeding in the immediate past and current feeding and a stronger negative association for feeding episodes in the early morning hours.